Nanothermodynamics: stable thermal equilibrium and nanoscale fluctuations

Ralph V. Chamberlin

Department of Physics, Arizona State University, Tempe, AZ, USA 85287-1504

*E-mail address:* ralph.chamberlin@asu.edu

**Abstract:** Nanothermodynamics describes the process where large systems subdivide into equilibrium distributions of small subsystems. A key ingredient is Hill's subdivision potential ($\mathcal{E}$) that ensures adherence to the 1st and 2nd laws of thermodynamics in systems of any size. In this review and reassessment, it is emphasized that nanothermodynamics gives new insight into many measurements, theories, and simulations. Measurements establishing the need for nanothermodynamics show thermodynamic heterogeneity from multiple effective temperatures ($T_i$) inside most types of materials. One theoretical result that requires $\mathcal{E} = 0$ is the stable solution of Ising's original model for finite chains of interacting spins, a solution Ising could not have found 40 years before Hill's work. Another result is a novel solution to Gibbs' paradox that makes the entropy of the semiclassical ideal gas exactly extensive. Molecular dynamics simulations reveal how a standard fluctuation relation is modified when local degrees of freedom fluctuate faster than their coupling to the heat bath, consistent with the measured thermodynamic heterogeneity. Simulations of a Creutz-like model, comprised of Ising spins coupled to an explicit heat bath of Einstein oscillators, are used to study the 2nd law. It is found that maximizing the total entropy ($S_t$) requires an intrinsically irreversible step, providing a counterexample to the usual claim that statistical mechanics emerges from reversible dynamics. Furthermore, fluctuations of this model are best described by Einstein's reversal of Boltzmann's relation and the 2nd-law, not by recent fluctuation theorems.

## 1. Introduction

Thermodynamics and statistical mechanics provide two complementary approaches for understanding thermal processes in nature. Thermodynamics contains the fundamental physical laws governing what can or cannot occur, while statistical mechanics provides microscopic insight and quantitative tools for interpreting thermal behavior. Most physicists prefer the detailed pictures and predictive power of statistical mechanics. Indeed, Feynman has called Boltzmann's factor a "fundamental physical law" that is the "the summit of statistical mechanics" [1]. But then he lists various assumptions needed to justify this "law." Basically, for Boltzmann's factor to fully describe any system, the system must have essentially instantaneous coupling to an effectively infinite and homogeneous heat bath. However, these assumptions are often unmet, especially for small systems that have finite-size thermal effects, including independently fluctuating small subsystems inside bulk samples. Therefore, a more basic approach is to first ensure that the system obeys the laws of thermodynamics before applying the tools of statistical mechanics. In practice, conservation of energy in thermal physics is usually found from the fundamental equation of thermodynamics, which combines the first and second laws.

Three people have made pioneering contributions to the fundamental equation of thermodynamics: Rudolf Clausius in 1850, Willard Gibbs in 1876, and Terrell Hill in 1962 (see Figure 1) [2,3,4]. The first two names and their contributions are well known to essentially all scientists, whereas Hill's contributions are relatively unknown. To put this into perspective, recall that it is legendary how Gibbs' work was underappreciated for 15-20 years, until it was translated into German and used with great success by early physical chemists. It has now been 60-65 years since Hill introduced a mathematically similar (and comparably important!) contribution to thermodynamics, and yet it is still relatively unknown. One explanation comes from Joel Keiser: "It may be, as with much of Terrell's work, that it was simply ahead of its time and that in future years much will be made of it," but this statement was made about 40 years ago [5]. Here I will clarify the importance of Hill's work by focusing on some key applications where statistical mechanics alone cannot explain the thermal behavior, so that one must go beyond Boltzmann's factor to find the stable thermal equilibrium and resulting nanoscale fluctuations.

The outline of this paper is as follows. **2) Background** reviews key concepts that aid in appreciating the power and utility of nanothermodynamics. **3) Measurements of Thermodynamic Heterogeneity** presents various experimental results that establish the need to use nanothermodynamics for most types of materials, including liquids, glasses, spin glasses, polymers, and crystals. **4) Stable Thermal Equilibrium of Standard Models** outlines the stable solution to Ising's original model for finite chains of binary spins, and for the semiclassical ideal gas that gives a novel solution to Gibbs' paradox. **5) Excess Energy Fluctuations in MD Simulations** describes how and why a standard fluctuation relation must be modified to connect standard statistical mechanics to many MD simulations. **6) Nanoscale Fluctuations and the Second Law of Thermodynamics** utilizes a simple model to test theoretical predictions for fluctuations of small systems, with the result that ideas based on thermodynamics yield better agreement with simulations than those based solely on statistical mechanics. **7) Summary** contains the main conclusions from this review.

## 2. Background

Clausius first codified the two basic laws of thermodynamics: 1) that energy is always conserved and 2) that some of this energy is degraded by entropy. Specifically, in 1850, he recognized that the single internal energy of a system can be changed by two distinct processes involving either heat or work. Then he recognized that entropy limits the ability of internal energy to do external work. Combining these two laws yields Figure 1A, an early version of the fundamental equation of thermodynamics, written in modern form. The left side of the equation contains the change in total internal energy, $dE_t$, while the right side of Figure 1A contains two pairs of conjugate variables from two distinct processes that can change $E_t$. The heat term comes from temperature times the change in total entropy ($TdS_t$), while the work term comes from pressure times the change in total volume ($pdV_t$) with the minus sign from work done on the environment.

A: $dE_t = TdS_t - pdV_t$ Clausius (1850)

B: $dE_t = TdS_t - pdV_t + \mu dN_t$ Gibbs (1876)

C: $dE_t = TdS_t - pdV_t + \mu dN_t + \mathcal{E} d\eta$ Hill (1962)

Figure 1. Key contributors to the fundamental equation of thermodynamics, with the pair(s) of conjugate variables that they added to ensure that energy can be conserved and entropy maximized.

Gibbs' greatest contributions to thermodynamics began in 1876 when he introduced a new pair of conjugate variables, the chemical potential times the change in total number of particles, $\mu dN_t$. The resulting fundamental equation for a simple system, such as a single-component ideal gas, is shown in Figure 1B. Gibbs also considered multiple types of particles, allowing the study of thermal equilibrium between different substances and between different phases of the same substance. This is the usual form of the fundamental equation found in textbooks. In general, the terms in Figure 1B apply only to the total quantities in effectively infinite systems, with an intensive variable ($T$, $p$, or $\mu$) times a change in an extensive variable ($dS_t$, $dV_t$, or $dN_t$). Because the equation in Figure 1B cannot systematically accommodate finite-size effects, non-extensive contributions to conservation of energy are missing from textbook treatments of thermodynamics and are similarly absent in standard statistical mechanics.

In 1962, Hill introduced the theory of small system thermodynamics [2]. As with Gibbs' contribution, the foundation of Hill's work involves a new pair of conjugate variables, the subdivision potential, $\mathcal{E}$, and the change in the number of subdivisions, $d\eta$. The resulting fundamental equation of thermodynamics is given in Figure 1C. In general, the number of small subsystems that are in a large system is given by $\eta + 1$. Thus, $\eta$ varies inversely proportional to the average number of particles in each subsystem, $\eta \approx N_t/\bar{n}$, in contrast to the total number of particles in the system, $N_t$, that is purely extensive. Therefore, while Gibbs' term is crucial for changes in large systems, $dN_t$, Hill's term is essential for small systems, $|d\eta| \sim |d\bar{n}/\bar{n}^2|$ with $\bar{n} \ll N_t$. Indeed, Hill's subdivision potential can be understood by comparison to Gibbs' chemical potential. Gibbs' $\mu$ is the change in (free) energy to take one particle from a bath of particles into the system, whereas Hill's $\mathcal{E}$ is related to the change in (free) energy to take a cluster of interacting particles from a bath of clusters into the system. In general, a cluster of $n$ interacting particles does not have the same energy as $n$ isolated particles. Specifically surface energies, length-scale effects, and fluctuations all contribute to the energy of a finite cluster of interacting particles, so that Hill's fundamental equation (Figure 1C) is required to systematically include these contributions for conservation of energy on all size scales. I will focus on small systems, especially small subsystems inside larger systems, where these contributions are difficult, if not impossible to include systematically in Boltzmann's factor. In such examples, Hill's small system thermodynamics is essential for understanding the stable thermal equilibrium and nanoscale fluctuations.

Hill developed his theory by considering large ensembles of small systems. Indeed, the foundation of his theory is that a large ensemble of small systems must behave like a large system, obeying the laws of standard thermodynamics. Starting in 1999, I adapted Hill's theory by imagining the mathematically similar but conceptually inverse picture of a large system that subdivides into an ensemble of small subsystems [6]. In 2000, I first published the term "nanothermodynamics" for this inverse picture [7]. In 2001 Hill published an explanation that he used the term "nanothermodynamics" as "a shortened and more fashionable version of 'small system thermodynamics'" [8]. Hill never fully accepted the inverse picture that I was working on, primarily because he could not imagine a mechanism for making an ensemble of independent subsystems inside a large system, as required for his theory. However, I stubbornly ignored his advice and continued to develop the inverse picture based on an abundance of experimental evidence showing thermodynamic heterogeneity inside bulk samples. One such experiment involves time-resolved specific heat (TRSH) measurements, especially those made by Meissner and collaborators [9].

Another experiment, that is essentially the inverse of TRSH, is non-resonant spectral hole burning (NSHB). NSHB was developed in collaboration with others starting in 1996 [10-12]. These experiments provide some of the earliest and most direct evidence for independently fluctuating subsystems inside bulk samples. Now this thermodynamic heterogeneity is well established in a wide variety of systems, including liquids, glasses, spin glasses, polymers, and crystals. Several examples are given in **Section 3**, below.

Hill's small system thermodynamics provides the theoretical foundation for the inverse process of nanothermodynamics, where a large system subdivides into a stable distribution of independent subsystems. This stable distribution is defined by setting Hill's subdivision potential to zero, $\mathcal{E} = 0$, as described in section 10-3 of Hill's 1964 book [3]. Indeed, quoting from this section: "Although $\mathcal{E}$ is negligible for a macroscopic system (. . .), it is not equal to zero in the strict sense that we are using $\mathcal{E} = 0$ above as an equilibrium condition. The macroscopic state is therefore not to be confused with the equilibrium state." In other words, macroscopic systems that are constrained to be homogeneous so that they can be treated using standard thermodynamics are in an unstable equilibrium, having $\mathcal{E} \approx 0$ but not $\mathcal{E} = 0$. Similarly, small systems with $\mathcal{E} \neq 0$ must also be constrained to prevent them from reaching their stable thermal equilibrium. With few exceptions, since 1964 no other researchers have used this $\mathcal{E} = 0$ stability condition for the treatment of small systems or subsystems, not even Hill himself. Therefore, to emphasize the importance of the stable thermal equilibrium, and to reduce confusion [13] with the work of others, I will continue to restrict use of the term "nanothermodynamics" to the inverse problem that requires $\mathcal{E} = 0$ for the stable equilibrium of independent subsystems inside larger systems, with "small system thermodynamics" used for the theory of small systems that are constrained to have $\mathcal{E} \neq 0$. It is difficult for me (as it was for Hill) to imagine how the inverse process of subdividing large systems could have $\mathcal{E} \neq 0$ for their internal subsystems.

Since the inception of nanothermodynamics as a way to understand how large systems subdivide into stable distributions of independent subsystems [6,7], a key goal has been to find applications that justify this extension of Hill's theory to bulk materials. At first the applications were postulated, based primarily on experimental evidence for thermodynamic heterogeneity. Some examples include a mesoscopic mean-field theory for supercooled liquids and the glass transition [6,14], a mean-field cluster model for non-classical critical scaling [7], and maximum entropy as a mechanism for $1/f$-like noise [15]. However, over the past decade, several theoretical results and computer simulations have clarified the connections. Therefore, an emphasis in this review will be the need to go beyond standard thermodynamics and statistical mechanics to include finite-size thermal effects in experiments, theory and simulations.

It is useful to preview some key results from the rest of this paper. **Section 3** focuses on experiments that show thermodynamic heterogeneity, providing direct evidence for the necessity of nanothermodynamics. **Section 4** describes how nanothermodynamics is needed to obtain the stable solutions of two standard models. First is the stable equilibrium of Ising's original (1925) model for finite chains of interacting binary spins, an equilibrium that Ising could not have found 40 years before Hill's work. Second is a novel solution to Gibbs' paradox, removing the residual non-extensive contributions to entropy that cause inconsistencies between statistical mechanics and thermodynamics. It is interesting to note that this solution to Gibbs' paradox is contained in the equations Hill published in his 1964 book [3], but at that time he seems not to have recognized the physical significance. The next two sections present results from computer simulations used to study nanoscale thermal fluctuations. **Section 5** describes molecular dynamics (MD) simulations that address Boltzmann's quest to find a connection between classical mechanics and statistical mechanics. The answer is that MD simulations give good agreement with statistical theories for thermal averages, but not for fluctuations. Instead, the MD simulations are quantitatively characterized by a theory that combines statistical mechanics with adiabatic modulations that are effectively decoupled from the heat bath, consistent with measurements of thermodynamic heterogeneity. **Section 6** utilizes a one-dimensional (1-D) Creutz-like model [16,17] for tests of the 2nd law of thermodynamics. An appealing feature of this model is that it has an explicit heat bath that allows calculation of the exact total entropy ($S_t$) of the system and its bath. One result is that an intrinsically irreversible microscopic step is needed to maximize $S_t$, providing a clear counterexample to the claim that the 2nd law is an emergent property of reversible dynamics. Another result that is a consequence of strict adherence to the 2nd law is that fluctuations in $S_t$ deviate from standard theories of thermal fluctuations based on statistical mechanics, including detailed fluctuation theorems [18-21]. Instead, the fluctuations are consistent with Einstein's reversal of Boltzmann's relation and the 2nd law. Speculations are made about the relevance of such models to real systems, and what might cause the intrinsically irreversible dynamics needed for thermal-equilibrium behavior. Finally, **Section 7** gives a summary and some conclusions.

## 3. Measurements of Thermodynamic Heterogeneity

### *3.1 Relevance of Heterogeneity to Nanothermodynamics*

If nanothermodynamics is to apply to large systems, there must be thermodynamic heterogeneity from an ensemble of independent subsystems inside the system. Similarly, any system that shows thermodynamic heterogeneity cannot be described by standard thermodynamics that applies only to systems having a heat bath that is homogeneous and effectively infinite. In other words, nanothermodynamics is irrelevant for any system that is large and homogeneous, but it is essential for systems of any size that exhibit thermodynamic heterogeneity. The purpose of this section is to outline some of the experimental evidence for thermodynamic heterogeneity that is found in virtually all types of materials, including liquids, glasses, spin glasses, polymers, and crystals.

Dynamic heterogeneity, where separate responses come from independent subsystems, has been a theoretical concept at least since 1874 when Boltzmann introduced his superposition principle [22]. Thermodynamic heterogeneity is closely related. In fact, nanothermodynamics applies to any system having internal fluctuations that are statistically independent. Such statistical independence is needed so that neighboring subsystems have uncorrelated dynamics, yielding entropies that are additive for an ensemble of independent subsystems inside a bulk sample. Because local entropies are difficult to measure directly, effective local temperatures ($T_i$) usually provide the most direct evidence for thermodynamic heterogeneity. Therefore, we describe measurements that imply distinct values of $T_i$ for independent degrees of freedom inside individual samples.

### *3.2 Time-Resolved Specific Heat Measurements*

Some of the earliest evidence for thermodynamic heterogeneity comes from time-resolved specific heat (TRSH) measurements at low temperatures, $T \ll 100$ K. The heterogeneity is deduced from time-dependent changes in effective temperature ($\Delta T_i$). Anomalies in $\Delta T_i$ reveal deviations from Debye's theory for the specific heat, $C = C_D T^3$, where Debye's theory yields values of $C_D$ from the elastic constants for harmonic modes. Specific heats that far exceed $C_D T^3$ were first found in amorphous materials [23,24] and attributed to defects [25,26]. However, ultra-pure single crystals also exhibit deviations from $C_D T^3$ [9]. Although the excess specific heat in crystals has been attributed to residual imperfections, the discovery of analogous excess energy fluctuations in MD simulations of crystals with no defects suggests that the excess specific heat may be intrinsic [27,28]. Evidence for excess specific heat is found in all materials that have been measured by TRSH at low enough temperatures. Symbols in Figure 2 show results from TRSH measurements on five different high-purity single crystals, given in the legend. The sketch in the insert helps to interpret the experiment. Each measurement is made by applying a heat pulse to one side of the sample and measuring the temperature as a function of time on the other side. Symbols in the insert show measurements from a quartz crystal at a base temperature of $T{\sim}0.2$ K. Note how $\Delta T_i$ starts at zero, rises sharply until it reaches a peak at $t{\sim}10$ μs, then recovers towards zero in multiple steps. The peak at ~10 μs is due to the arrival of ballistic phonons from the heater to the thermometer. After the peak, as the added heat becomes dispersed throughout the heat bath, $\Delta T_i$ stabilizes at a short-time plateau that is consistent with $C_D T^3$ for this crystal (black line). However, starting at $t{\sim}1$ ms, $\Delta T_i$ relaxes again, reaching a long-time plateau (red dashed line) at $t > 10$ ms that gives the equilibrium specific heat for the sample. Finally, at still longer times ($t > 0.5$ s, not shown) $\Delta T_i$ returns to zero as the excess heat flows out of the fine wires that connect the sample to the cryostat.

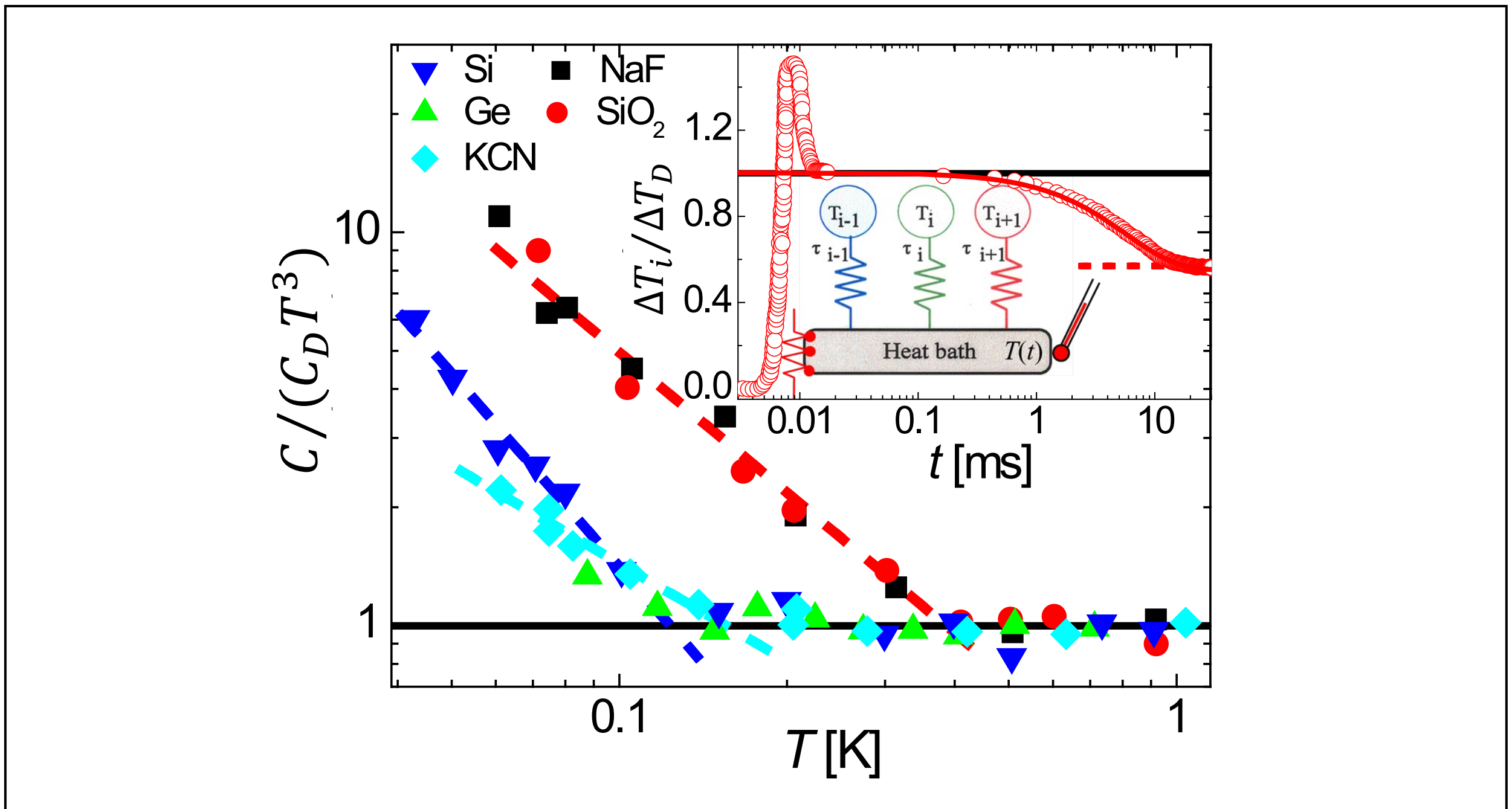

*Figure 2* Solid symbols show specific heat of five high-purity single crystals, listed in the legend, from measurements of temperature as a function of time (insert). (Data are from [9], with presentation from [28].) Sketch in the insert shows "box" model for analyzing experiments. Each box (set of localized degrees of freedom) has a local temperature ($T_i$) and heat resistor (time constant, $\tau_i$) coupling it weakly to the heat bath. A heat pulse is applied to one side of the sample with temperature measured as a function of time on the other side. Insert graph shows relative change in the $T_i$ as a function of time in the quartz crystal ($SiO_2$) at ~0.2 K, with the short-time (Debye like) behavior shown by the solid (black) line and long-time (equilibrium) behavior shown by the dashed (red) line. An exponential fit to the data gives the solid red line. Main figure shows equilibrium (long-time) values of specific heat as a function of temperature, normalized by the average of the initial (short-time) values that show Debye-like ($C_D T^3$) behavior (solid line at 1). Dashed lines show power-law fits to the NaF (black) and Si (blue) data.

Note that specific heat is found from the inverse of $\Delta T_i$. Thus, the insert in Figure 2 shows that the equilibrium specific heat (red dashed line) is significantly higher than that predicted by Debye's theory (black line). Furthermore, during the nearly exponential relaxation that occurs between 1 ms<t<10 ms the excess heat supplied to the sample is not flowing across the sample (which happens at t<0.1 ms), and it is not flowing out of the sample (which happens at t>100 ms), so it must be flowing into distinct degrees of freedom inside the sample. Specifically, the first plateau (that agrees with Debye's theory) comes from the homogeneous heat bath of phonons, while the second plateau is due to some other degrees of freedom in the sample. From the zeroth law of thermodynamics, net heat flows only if there is a temperature difference. Thus, this experiment establishes that there are multiple temperatures inside a sample, which persist for times $t \gg 1$ ms. Furthermore, while these slow degrees of freedom are relaxing (or fluctuating), their exchange of energy with the heat bath is slow ($t \gg 1$ ms), thereby failing the fundamental requirement of fast thermalization needed for Boltzmann's factor. This is direct evidence of thermodynamic heterogeneity. For such slow degrees of freedom, temperature is not a good thermodynamic variable, so that a full understanding requires going beyond Boltzmann's factor to describe the fluctuations and relaxation. When measured by TRSH at low enough temperatures, similar slow relaxation is found in every sample studied, including dielectric crystals and amorphous materials.

Solid symbols in Figure 2 show the temperature dependence of the long-time (equilibrium) values of normalized specific heat from TRSH measurements of five different high-purity single crystals, given in the legend. The solid black line at $C_p/C_D T^3 = 1$ shows the $T$-dependence expected from Debye's theory. The dashed lines come from power-law fits to the low-$T$ data for NaF (black squares) and Si (blue triangles), showing how the equilibrium specific heat diverges from Debye behavior with an inverse power of $T$. This excess specific heat, that has been attributed to residual defects that are unavoidable in even ultra-pure crystals, could instead be due to excitations that are localized by anharmonic interactions. The anharmonicity decouples the interactions from the heat bath of phonon modes, yielding excitations that conserve local energy instead of having a single well-defined $T$. Qualitatively similar deviations at low temperatures are found in energy fluctuations of MD simulations, as described in Section 5, below.

TRSH measurements establish that there are at least two distinct temperatures in most (if not all) samples during their slow response that yields the excess specific heat at low temperatures. At higher temperatures, the excess specific heat becomes much smaller than standard Debye behavior. Specifically, extrapolating the dashed line for NaF in Figure 2 to 1 K, the excess contribution is about 30% of the Debye value, and by room temperature the excess will be in the ppm range. However, at all temperatures, the excess specific heat gives 100% of the slow response that leads to thermal equilibrium. Non-resonant spectral hole burning (NSHB) provides additional information about the slow degrees of freedom that cause the thermodynamic heterogeneity that justifies the need for nanothermodynamics.

*3.3 Nonresonant Spectral Hole Burning in a Spin Glass*

The sketch in the upper right of Figure 3 depicts the so called "box model" that is used to characterize the behavior of NSHB [12]. Thus, NSHB is essentially the inverse of TRSH (Figure 2). Specifically, in NSHB, a large amplitude low-frequency oscillation in an external field adds excess energy to slow degrees of freedom that are selected by the frequency ($f$) of the field. This oscillation can be in the electric field for dielectrics [10], in the magnetic field for magnetic materials [11], or in the stress for mechanical response [29]. The resulting excess energy changes the time scale of the response, which can be interpreted as a change in local temperature, $\Delta T_i$, due to "heating" of selected degrees of freedom in the sample. Usually, the slow degrees of freedom couple weakly to the heat bath (heat resistors in the sketch), yielding the set of slow relaxation times ($\tau_i$) for the $\Delta T_i$ to return to zero.

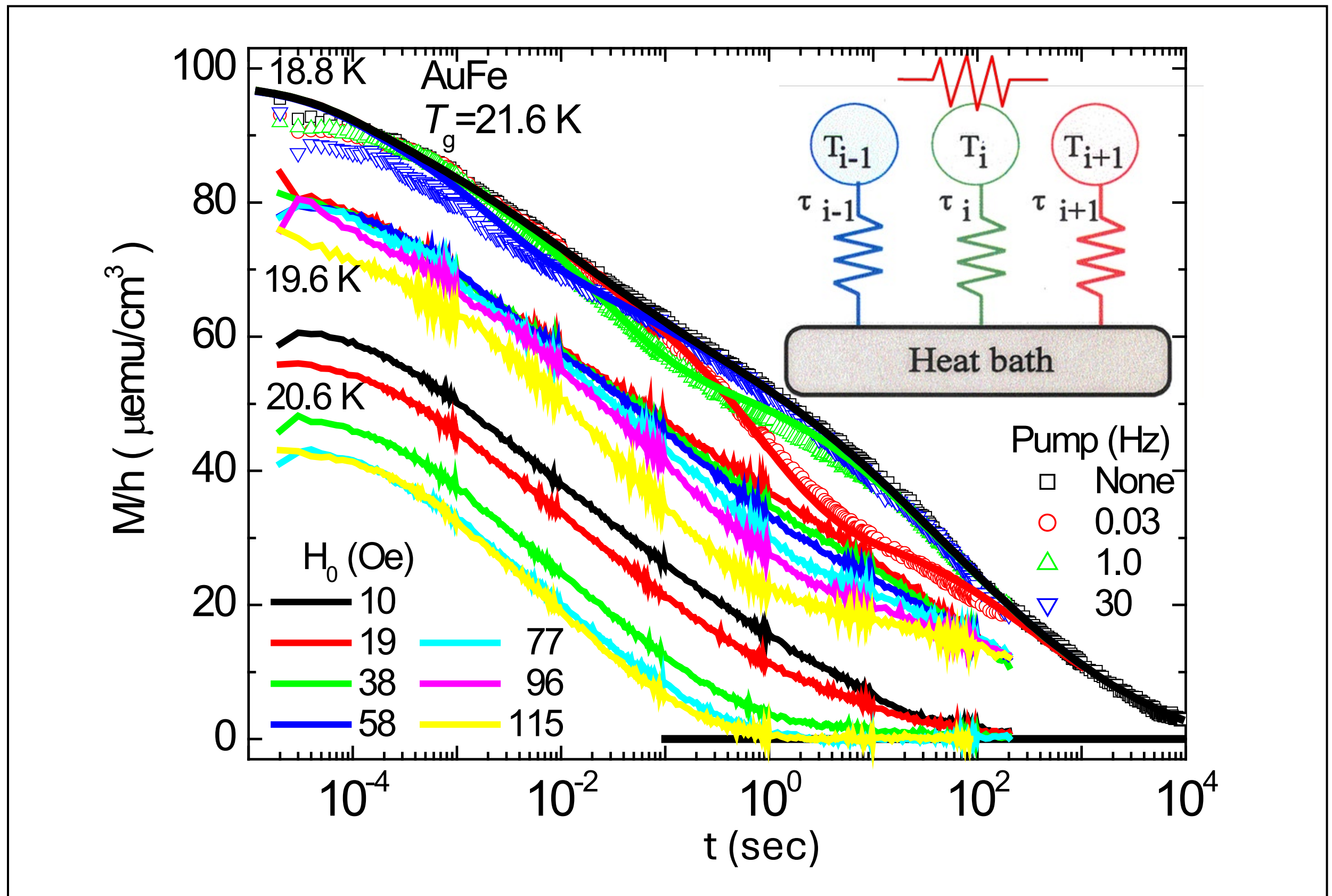

*Figure 3* Magnetization as a function of time from a 5% Au:Fe spin glass sample after pump oscillations of magnitude $0 \leq H_0 \leq 115$ Oe and frequency $0.03 \leq f \leq 30$ Hz. (Adapted with permission from Ref. [11]. Copyrighted by the American Physical Society.) Spectral holes at $T = 18.8$ K show negligible overlap between degrees of freedom modified by $f = 30$ Hz (blue) and 0.03 Hz (red) establishing that the nonlinear response involves thermodynamic heterogeneity. The solid black curve is a fit to the unpumped response from a distribution of relaxation times that yield the $\tau_i$ for the box model (insert) that is used to interpret the data, similar to the model in Fig. 2 except that excess energy is added directly to slow degrees of freedom selected by their absorption of energy at frequency, $f$. Other solid curves show predictions of the model with the specific heat of the spins as the only adjustable parameter. Spectral steps (where $M/h$ does not overlap at short times) at $T = 19.6$ and 20.6 come from degrees of freedom that have a spin temperature above the glass temperature ($T_g = 21.6$ K), while the rest of the sample remains a spin glass, establishing that the nonlinear response can be interpreted as a change in effective local (spin) temperature.

The main part of Figure 3 shows NSHB measurements on a 5% Au:Fe spin glass [11,12]. This sample has a sharp peak in the field-cooled magnetization (not shown) that defines a spin-glass temperature of $T_g = 21.6$ K. NSHB measurements are made utilizing a single oscillation "pump" in the magnetic field having a large-amplitude ($H_0$) at low $f$, followed by a small step in the field (*h*=8 Oe) to measure the linear response of the sample after being nonlinearly modified by the pump. The uppermost set of data in Figure 3 are from NSBH measurements at $T = 18.8$ K. Symbols show the linear response with no pump (black squares), and after a pump oscillation of $H_0 = 96$ Oe at $f = 0.03$ Hz (red circles), 1.0 Hz (green up triangles), and 30 Hz (blue down triangles). Even in the raw data the influence of the pump frequency is clear: $f = 0.03$ Hz shifts the slow degrees of freedom at long times, $f = 30$ Hz shifts the "fast" degrees of freedom at short times, while $f = 1.0$ Hz shifts in the middle. Indeed, there is virtually no overlap in the degrees of freedom shifted by the lowest- and highest-frequency pumps, clearly and directly establishing the heterogeneous nature of the non-exponential response. In other words, there is negligible correlation between the slow and fast degrees of freedom. Solid curves show predictions of the box model with the distribution of relaxation times defined by the linear response (black), and hole burning behavior (red, green, blue) found from the known pump frequency and amplitude. The model gives a specific heat per spin of $\Delta C/k = 5x10^{-7}$ as the only free parameter.

Such a tiny specific heat per spin implies that relatively small magnetic fields can greatly increase the spin temperature above the bath temperature, consistent with the behavior shown in Figure 3 for NSHB at $T = 19.6$ and at 20.6 K. First, at $T = 19.6$ K, note that pump oscillations of $H_0 \leq 96$ Oe still produce spectral holes due to simply shifting the response times with no net loss in amplitude over the measurement time window. However, for $H_0 = 115$ Oe, a spectral "step" emerges where significant amplitude is lost from the time window. This spectral step indicates that many degrees of freedom are "heated" by at least 2 K to above $T_g$, where they respond on paramagnetic time scales ($t \gg 1\mu s$), while the rest of the sample remains a spin glass. This interpretation is confirmed by measurements at $T =$ 20.6 K, where $H_0 \geq 19$ Oe is sufficient to heat a significant fraction of the degrees of freedom into the paramagnetic phase. Measurements, modeling, and interpretations from Figure 3 indicate that spin temperature can differ greatly from bath temperature during response measurements. Such significant changes in spin temperature are well known in magnetism [30], including spin temperatures in magnetic resonance that can be negative. Thus, local temperatures that differ from the bath, with multiple values that depend on position in the sample are useful for interpreting the behavior, providing evidence for the thermodynamic heterogeneity that is needed for nanothermodynamics.

*3.4 Nonresonant Spectral Hole Burning in Glass-Forming Liquids*

Figure 4B shows dielectric NSHB measurements on a glass-forming liquid (propylene carbonate, PC) at $T = 187.5$ K, slightly above the glass temperature $T_g \approx 186$ K [10,12]. Again, there is negligible overlap in the degrees of freedom modified by the highest and lowest frequencies, directly establishing the heterogeneous nature of the slow response. For PC, the specific heat per molecule from the slow response is large enough to be measured directly, $\Delta C/k \approx 7.85$ [31]. For theoretical analysis, knowing $\Delta C$ has the advantage that the box model (Figure 3) can be used with no adjustable parameters. Results utilizing a simplified box model (without heat resistors, $\tau_i \to \infty$) are shown by the smooth curves that mimic the measurements. Better agreement with the data, especially after the pump oscillation at lowest frequency, can be obtained (also with no free parameters) by using heat resistors with a distribution of recovery times ($\tau_i$) that match the distribution of relaxation times in the equilibrium response [32]. Each $\tau_i$ governs how the associated effective local temperature returns to the bath temperature, $\Delta T_i \to 0$. This model quantifies what must happen for systems that obey Joule heating, Ohm's law, and Fourier's law. The main experimental challenge from the large $\Delta C$ is that very high fields are needed to yield measurable spectral holes, so that it is difficult to increase any effective local temperature to more than ~0.04% above the bath temperature. Still, the axes in Figure 4B show how the shift in response times after the pump oscillation (left axis) can be converted to a change in effective local temperatures (right axis). Again, it is useful to characterize the slow response using multiple internal temperatures, yielding the thermodynamic heterogeneity needed for nanothermodynamics.

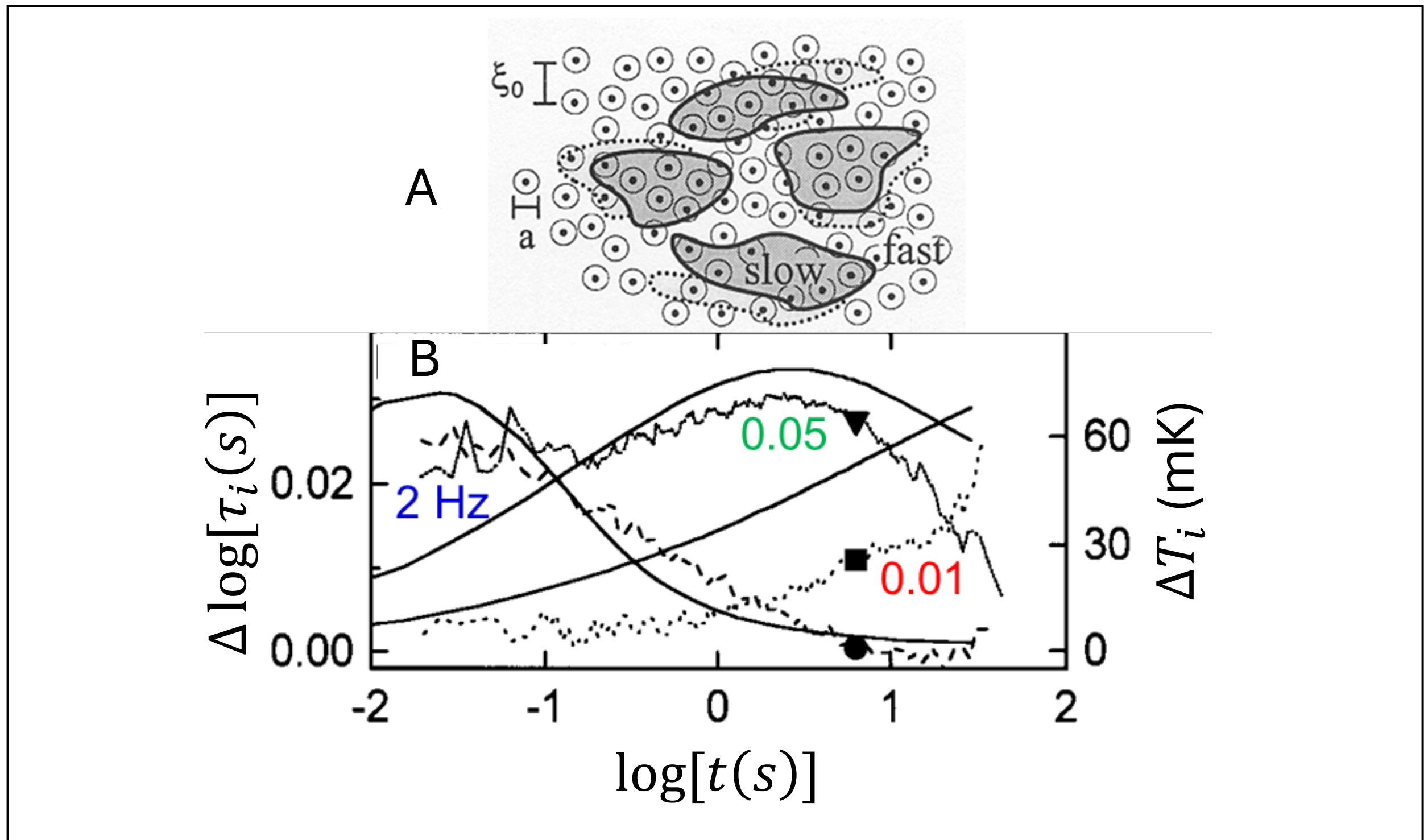

*Figure 4*. (A) Sketch showing fast and slow regions, deduced from multidimensional NMR measurements that yield length scales of 1-3 nm for the dynamic heterogeneity, from [33] (used with permission). (B) NSHB measurements on propylene carbonate at $T = 157.4$ K after pump oscillations at the three frequencies shown. The change in response time (left axis) can be converted to a change in effective local temperature (right axis) as a function of time. The solid curves come from a simplified box model, with no adjustable parameters. From [10] (used with permission).

Although NSHB and TRSH establish the ubiquity of thermodynamic heterogeneity in a wide variety of substances, other techniques are needed to determine the length scale of the heterogeneity. Perhaps the least invasive and most direct technique involves multi-dimensional nuclear magnetic resonance (NMR). Figure 4A is a picture that emerges from these measurements [33]. Analysis of the measurements yields typical length scales of 1-3 nm [34] (~10-100 molecules) for the thermal and dynamic heterogeneity of slow degrees of freedom in glass-forming liquids. Such nanoscale lengths and distributions of independent fluctuations inside bulk samples emphasize the need for nanothermodynamics.

## 4. Stable Thermal Equilibrium of Two Standard Models

### *4.1 Relevance to Nanothermodynamics*

Perhaps the most important application of nanothermodynamics is to treat finite size thermal effects in finding the stable equilibrium of systems, especially when a large system subdivides into subsystems. Subdividing into internal subsystems is usually favored due to an increase in entropy that lowers the net free energy. For sufficiently simple models, Hill's subdivision potential can be calculated directly, allowing Hill's fundamental equation to be used to determine the stable thermal equilibrium. Two such simple systems are Ising's original model for magnetic spins, and the semiclassical ideal gas.

### *4.2 Stable Thermal Equilibrium of Ising's Model for Finite Chains of Interacting Spins*

Detailed histories of Ising and his model are given in [35,36]. Briefly, in 1925 Ernst Ising published his solution to the Lenz-Ising model for finite chains of interacting binary degrees of freedom ("spins") [37]. His solution involved summing over all possible chain lengths, then forcing the chain to be infinitely long and homogeneous. Since 1925, most (but not all [38-40]) discussions of the Ising model have also assumed that the spin system is infinite and homogeneous, but this constraint prohibits the system from finding its true equilibrium if finite-size effects are included. Here, we focus on the recent discovery of the stable thermal equilibrium to Ising's original model [13,39]. Although Ising could not have found this equilibrium 40 years before Hill's work, it is now a useful exercise that elucidates the necessity of nanothermodynamics.

Most modern treatments of the Ising model start by assuming that the system is infinite and homogenous, as represented by the upper sketch in Figure 5. This limit is needed for standard thermodynamics. However, in Ising's original work he treated finite chains, shown by the right-side sketch in Figure 5, which require nanothermodynamics. This right-side sketch shows several important features. Note that spin interactions are low energy between aligned spins (•), high energy between antialigned spins (X), or no energy when there is a break in the interaction (**o**). Breaks in the interaction subdivide the initially infinite and homogeneous system into finite chains, as originally treated by Ising. Also shown with the spins are two types of heat baths. The upper bath is idealized, assumed to be effectively infinite and homogeneous with essentially instantaneous coupling to every spin, thereby directly imposing a uniform temperature on the spins. The lower bath is explicit, comprised of a set of Einstein oscillators (Creutz "demons"), which accommodates heterogeneity in the spin system and facilitates testing the validity of Boltzmann statistics. For example, local temperatures can be extracted from the occupation probabilities in the equally spaced energy levels. Furthermore, this explicit heat bath provides information about the spins as they thermalize and allows the spins and bath to form a closed system for testing the 2nd law of thermodynamics (see Section 6, below).

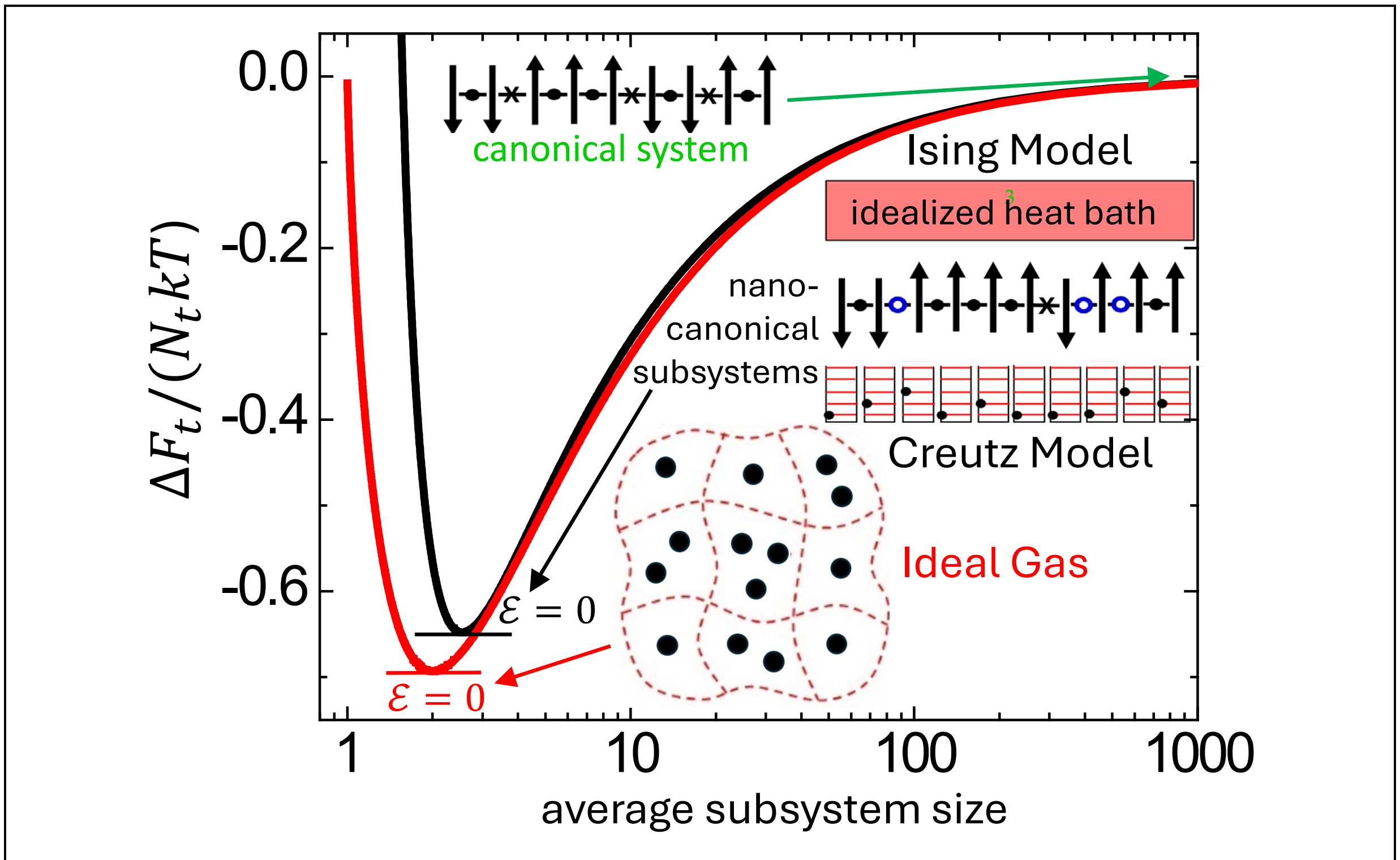

*Figure 5* Change in total free energy ($\Delta F_t$), normalized by the total number of particles in the system ($N_t$), as a function of the average number of particles in the subsystems ($\bar{n}+1$ or $\bar{N}$). Lines show behavior of the 1-D Ising model using $KT/J = 1$ (black) and semiclassical ideal gas (red). The lower sketch depicts a semi-classical ideal gas that is in the nanocanonical ensemble, having an equilibrum distribution of small subsystems. The upper sketch depicts the 1-D Ising model in the canonical ensemble, constrained to be large and homogeneous, $\bar{n}+1 \rightarrow N_t$. Total free energy is reduced by subdividing because entropy increases when there are no external constraints on subsystem size. Right-side sketch depicts interactions between spins that may be low energy (•), high energy (x), or no energy (o) due to breaks in the interaction. These breaks yield finite chains, as in Ising's original model. Stable equilibrium occurs at the minimum in total free energy, $\mathcal{E} = 0$, which defines the nanocanonical ensemble and yields $\bar{n}+1 = 1 + cosh(J/kT)$ for the Ising model. Spin systems studied have two types of heat baths: idealized (upper) that is assumed to be infinite and homogeneous, and explicit (lower) from a set of Einstein oscillators (Creutz "demons").

The Lenz-Ising Hamiltonian for a 1-D chain of $n$ links ($n+1$ spins) in zero field is

$$U(n, n_x) = -J \sum_{i=0}^{n} \sigma_i \sigma_{i+1}. \quad (1)$$

Here, $J$ is the interaction energy between nearest-neighbor spins, with $\sigma_i = +1$ if the $i^{th}$ spin is up and $\sigma_i = -1$ if it is down. (Note that volume, $V$, is absent from Eq. (1) because the energy is assumed to be independent of density.) Referring to the sketches of spins in Figure 5, this energy can be expressed in terms of the number of high-energy interactions ($n_x$) by

$$U(n, n_x) = -J(n - 2n_x). \tag{2}$$

First assume that the spin chain is ideally coupled to a heat bath having temperature $T$ so that Boltzmann's factor applies to the relative probability of each energy. Using the binomial coefficient for the multiplicity of these energies, the partition function for an Ising chain having a fixed number of spins ($n + 1$) comes from summing over all energies

$$Q(n, T) = \sum_{n_x=0}^{n} \frac{2n!}{(n-n_x)!n_x!} e^{-U(n,n_x)/kT} = 2^{n+1}[\cosh(J/kT)]^n \tag{3}$$

Equation (3) is the canonical-ensemble partition function. It is consistent with the zero-field limit of the partition function that Ising gave in the middle of pg. 256 in [37]. Next assume that the spin chain is ideally coupled to a bath of spins having chemical potential $\mu$ with $e^{\mu/kT} < 1/[2\cosh(J/kT)]$ to ensure convergence. Because there are $n + 1$ spins in the chain the chemical potential term is raised to the power of $n + 1$. Summing the resulting series gives:

$$\Upsilon(\mu, T) = \sum_{n=0}^{\infty} Q(n, T)\, e^{\mu(n+1)/kT} = 2e^{\mu/kT}/\left[1 - 2e^{\mu/kT}\cosh(J/kT)\right]. \tag{4}$$

Equation (4) is the generalized-ensemble partition function. Although Eq. (4) is formally equivalent to Eq. (7) in [37], Ising used the dummy variable $x \leftrightarrow e^{\mu/kT}$ as a tool to study the mathematics of 1-D chains. Ising's mathematical treatment made sense in 1925 because Eq. (4) (with no extensive environmental variables to fix the size of the system) is ill-defined in standard statistical mechanics. Small-system thermodynamics is needed to justify the physics of Eq. (4), which gives Hill's subdivision potential

$$\mathcal{E} = -kT\ln(\Upsilon). \tag{5}$$

The average number of spins in a chain is

$$\bar{n}(\mu, T) + 1 = \frac{\partial \ln(\Upsilon)}{\partial(\mu/kT)} = \left[1 - 2e^{\mu/kT}\cosh(J/kT)\right]^{-1} \tag{6}$$

Because Eqs. (4-6) are in the generalized ensemble, standard thermodynamics and statistical mechanics give no guidance on how to proceed. A key question is: what value of $\mu/kT$ is needed for Eq. (6) to yield the equilibrium average chain length as a function of $T$? It is useful to contrast Eq. (6) with the usual procedures for obtaining $\mu/kT$ and $\bar{n}(\mu, V, T)$ in the grand-canonical ensemble. If there is a fixed density of particles, then $\mu/kT$ is chosen to give this $\bar{n}/V$. Familiar examples include the densities of charge carriers in conductors and bosons in superfluids. Alternatively, if the density of particles is not fixed, then $\mu/kT = 0$ is used to minimize the free energy when the system itself controls the density of particles. Familiar examples include photons in black-body radiation and phonons in solids.

Using Eq. (6) to write the chemical potential in terms of the average number of links in a chain, $e^{\mu/kT} = 1/[2(1 + 1/\bar{n}) \cosh(J/kT)]$, the subdivision potential becomes:

$$\mathcal{E} = -kT \ln(\Upsilon) = -kT \ln[\bar{n}/\cosh(J/kT)] \qquad (7)$$

In standard statistical mechanics, $\mu = 0$ must be used if there are no external constraints on the density of particles. In nanothermodynamics, $\mathcal{E} = 0$ must be used if there are no external constraints on subsystem size. Thus, the average number of spins per finite chain in the stable thermal equilibrium of Ising's original model is given by:

$$\bar{n} + 1 = 1 + \cosh(J/kT) \qquad (8)$$

The free energy reduction as a function of average subsystem size, black curve in Figure 5, gives insight into understanding the thermal stability. This $\Delta F_t$ is obtained from Hill's fundamental equation, Figure 1C, by integrating the subdivision potential over the number of subdivisions, $\Delta F_t = \int_0^{\bar{\eta}} \mathcal{E} d\eta$. For Figure 5, a change of variables is used to integrate over $n = N_t/(\eta + 1)$. Starting with an effectively infinite system, $N_t \to \infty$, the change in free energy per particle to subdivide the system into subsystems of average size $\bar{n}$ is:

$$\Delta F_t/(N_t kT) = \int_0^{\bar{\eta}} (\mathcal{E}/kT) d(\eta/N_t) = \int_{\infty}^{\bar{n}} \ln[n/\cosh(J/kT)]\, dn/n^2$$
$$= -\{\ln[\bar{n}/\cosh(J/kT)] + 1\}/\bar{n} \qquad (9)$$

The black curve in Figure 6 shows Eq. (9), using $J/kT = 1$ for specific values. The minimum free energy occurs at $\partial(\Delta F_t)/\partial\bar{n} = 0$, yielding $\mathcal{E} = 0$ and $\bar{n} + 1 = 1 + \cosh(J/kT) = 2.543$ .... The resulting reduction in total free energy due to subdividing is significant at all nonzero temperatures. Thus, the large homogeneous solution is never in thermal equilibrium at any temperature, $T > 0$. Indeed, the stable thermal equilibrium of Ising's original model cannot be found from standard thermodynamics or statistical mechanics alone, instead requiring insight from nanothermodynamics to obtain the minimum in free energy.

*4.3 Equivalence Between Ising's Original Model and a Model for Random Interactions*

Equation (8) gives the average number of spins for the Ising chain in stable thermal equilibrium. It is obtained from the 1-D Ising Hamiltonian [Eq. (1)] using the nanocanonical ensemble that is unique to nanothermodynamics. The same answer is obtained in the canonical ensemble from standard statistical mechanics by using a different Hamiltonian

$$U(N, N_x, N_0) = -\sum_{i=1}^{N} J_i \sigma_i \sigma_{i+1} \qquad J_i = (J, 0) \qquad (10)$$

Here, $N$ is the total number of links between $N + 1$ spins, with $N_x$ the number of high-energy links and $N_0$ the number of links having no interaction energy, $J = 0$. The right-side sketch in Figure 5 shows $N_0 = 3$. Equation (10) differs from Eq. (1) by having two values for the exchange interaction, $J$ or 0. If $J_i$ had a frozen distribution, Eq. (10) would be analogous to a 1-D Edwards-Anderson spin glass model with a bimodal distribution [41]. Instead, we allow the interactions to change freely, $J \leftrightarrow 0$, yielding a thermal distribution of $J_i$. Using the number of each type of link, the energy of Eq. (10) can be written as:

$$U(N, N_x, N_0) = -J(N - 2N_x - N_0). \quad (11)$$

Utilizing the trinomial coefficient for the multiplicity of each energy, the partition function for an Ising chain having a fixed number of spins $(N + 1)$ comes from summing over all energies

$$Q(N,T) = \sum_{N_0=0}^{N} \sum_{N_x=0}^{N-N_0} \frac{2^{N_0+1} N!}{(N-N_x-N_0)! N_x! N_0!} e^{-U(N,N_x,N_0)/kT} = 2^{N+1}[1 + \cosh(J/kT)]^N \quad (12)$$

This is the partition function for a 1-D random-exchange Ising model in the canonical ensemble. It gives an average energy of

$$\overline{U} = \frac{\partial \ln Q}{\partial(-1/kT)} = -NJ \frac{\sinh(J/kT)}{1+\cosh(J/kT)} \quad (13)$$

This $\overline{U}$ differs from the usual expression for the average energy of the Ising model because many links between spins do not contribute to the energy. However, using the average number of breaks in the interaction, $\overline{N_0} = N/[1 + \cosh(J/kT)]$ [13], the average energy per interacting link is identical to the standard Ising model

$$\overline{U}/(N - \overline{N_0}) = -J \tanh(J/kT). \quad (14)$$

Furthermore, using the probability of a break in the interaction, $p = \overline{N_0}/N$, with percolation theory in the thermodynamic limit, the average number of interacting links between breaks in the interaction is:

$$\bar{n} = \sum_{n=0}^{\infty} n(1-p)^n p^2 / \sum_{n=0}^{\infty} (1-p)^n p^2 = (1-p)/p = \cosh(J/kT) \quad (15)$$

equivalent to Eq. (8). This emphasizes another key difference between standard statistical mechanics and nanothermodynamics. If finite-size thermal effects are included, the behavior depends on the choice of ensemble. Moreover, different Hamiltonians solved in their appropriate ensembles can yield identical results, but only if the subsystems are in the stable thermal equilibrium from the nanocanonical ensemble of nanothermodynamics.

*4.4 Thermal Equilibrium of the Semi-Classical Ideal Gas*

Nanothermodynamics provides a novel solution to Gibbs' paradox that has several advantages over previous solutions [13,40]. First, the nanocanonical ensemble yields entropy that is exactly extensive for systems of all sizes, unlike previous solutions that are nonextensive for small systems. Specifically, entropy is precisely doubled when two systems having the same average number of subsystems are combined. Indeed, because previous solutions of Gibbs' paradox neglect finite-size effects, nanothermodynamics is the only way to obtain extensive entropy for small systems. Second, as depicted in the sketch at the bottom of Figure 5, particles are indistinguishable only if they are within the same subsystems, allowing distant particles in large systems to be distinguishable by their locations when in distinct subsystems. Third, nanothermodynamics provides an explanation for the empirical accuracy of the Sackur-Tetrode equation. Specifically, this nanothermodynamic solution may be necessary to explain the absence of isotope effects in the measured entropies of real gases [42], thereby solving various questions about the entropy of mixing [43]. Fourth, as with Ising's model for finite chains of spins, only the nanocanonical ensemble yields the minimum free energy of the semiclassical ideal gas. The remainder of this section outlines the derivation and advantages of the nanothermodynamic solution.

As with Ising's original model, Hill's subdivision potential can be calculated exactly for the semi-classical ideal gas in the generalized ensemble. In fact, Hill gives an answer in his 1964 book [3] (Eq. 10-88) $\mathcal{E} = -kT\ln(1+\overline{N})$, where $\overline{N}$ is the average number of particles in a subsystem. However, this result assumes that the sum over particles in each subsystem starts at $N = 0$, whereas Hill's stability condition requires that each subsystem has at least one particle. Indeed, the sentence associated with the first footnote in section 10-3 of [3] states that "an empty 'system'…is not counted as a system of the ensemble. This is the natural, realistic, and necessary point of view…in a solvent or gas." Therefore, starting the summation at $N = 1$ in Hill's Eq. 10-85 gives

$$\mathcal{E} = -kT\ln(\overline{N} - 1). \tag{16}$$

This term lowers the total free energy when a large system subdivides into subsystems. Most previous studies of small-system thermodynamics assume $\mathcal{E} \neq 0$, which applies only if there are external constraints on the average subsystem size [44]. Instead, $\overline{N} = 2$ is needed to yield $\mathcal{E} = 0$ for the stable equilibrium of nanothermodynamics. Any value $\overline{N} \neq 2$ fails to minimize the free energy, as indicated by the red curve in Figure 5. In fact, it is difficult to imagine how external constraints could control the distribution of subsystems inside bulk samples to yield $\mathcal{E} \neq 0$. An analogy from standard statistical mechanics would be if there were external constraints on a black body that controlled the distribution of photons by imposing $\mu/kT \neq 0$. Therefore, we focus on the stable equilibrium of nanothermodynamics.

The lower sketch in Figure 5 depicts how a semiclassical ideal gas can be subdivided into small subsystems. Ideal gases have no interactions, so that subsystems must be identified by another mechanism. We relate the dashed lines in the sketch with boundaries between subsystems of distinguishable particles. Specifically, all particles within a subsystem are indistinguishable from each other, but they are distinguishable from particles in other subsystems. Because distinguishable particles have higher entropy than indistinguishable particles, the total entropy of a semi-classical ideal gas is significantly increased by subdividing a large homogeneous system into a heterogeneous ensemble of small subsystems, which reduces the free energy. A similar concept also applies to Ising's model. In real systems, the interaction energy $J_i = J$ between neighboring spins usually involves quantum mechanical exchange that requires particles to be statistically indistinguishable. In this case, $J_i = 0$ would come from breaks in the interaction due to neighboring spins that are distinguishable, yielding the increase in net entropy that lowers the free energy and gives the stable thermal equilibrium.

As with Ising's model, it is instructive to calculate the change in free energy per particle due to subdividing the semiclassical ideal gas. Starting from the thermodynamic limit ($N_t \to \infty$) the answer is:

$$\Delta F_t/(N_t kT) = \int_0^{\bar{\eta}} (\mathcal{E}/kT)(d\eta/N_t) \; = -\int_{\infty}^{\bar{N}} (\mathcal{E}/kT) dN/N^2$$

$$= [(\bar{N}-1)\ln(\bar{N}-1) - \bar{N}\ln(\bar{N})]/\bar{N} \qquad (17)$$

The red line in Figure 5 shows Eq. (17) as a function of $\bar{N}$. The minimum at $\bar{N} = 2$ occurs when most particles are distinguishable by belonging to separate subsystems, which significantly increases the entropy and lowers the free energy. Using $\bar{N} = 2$, Eq. (17) gives $\Delta F_t/(N_t kT) = -\ln(2)$, yielding the net reduction in free energy per particle when a semiclassical ideal gas subdivides into its stable thermal equilibrium of subsystems.

## 5. Excess energy fluctuations in MD simulations

### *5.1 Equilibrium Averages and Fluctuations in Statistical Mechanics*

Boltzmann often sought to connect classical mechanics to statistical mechanics, with mixed success. Now, molecular dynamics (MD) simulations provide a powerful tool for investigating Boltzmann's dream that classical forces and Newton's laws may yield thermal behavior in complex systems. In fact, the first MD simulation on a digital computer tested whether anharmonic interactions between idealized atoms would yield average energies that obey the equipartition theorem characteristic of thermal equilibrium. Specifically, in 1953, Fermi, Pasta, Ulam and Tsingou (FPUT) used what was then the world's most powerful computer to simulate 1-D systems of up to 64 particles [45]. Although they knew that harmonic modes would not mix, they thought that adding anharmonic interactions would allow the system to evolve towards thermal equilibrium. Instead, they found that there was "very little, if any, tendency toward equipartition of energy among the degrees of freedom."

The failure of the FPUT model to reach thermal equilibrium has led to various interpretations. One view is that the FPUT model is too small and simple to yield the stochastic dynamics needed for thermalization, at least when simulated at low energies [46,47,48]. Indeed, MD simulations of large systems often show good agreement with the equipartition theorem expected for thermal equilibrium. However, when Fermi's idea of using MD simulations to study thermal averages is extended to study thermal dynamics, significant deviations are found from the standard fluctuation relation that is based on the equipartition theorem [27]. Thus, fluctuations are a more-stringent test of thermal behavior than equilibrium averages. One way to investigate fluctuations in small systems is to study large simulations that are subdivided into smaller subsystems or "blocks" [49,50,27]. Because there are interactions between the blocks, they cannot be treated as an ensemble of independent subsystems needed for Hill's theory. Nevertheless, some ideas from small-system thermodynamics may still apply, such as thermal behavior that depends on the choice of ensemble and type of dynamics.

A basic test of whether MD simulations yield standard statistical mechanics comes from the connection between energy fluctuations and the equipartition theorem via specific heat ($C = \partial E / \partial T$). The canonical ensemble has a formal identity that makes this connection, which we call the $C \leftrightarrow (\Delta E)^2$ relation. It is useful to review the derivation of this relation. The terminology is from Ref. [28]. Let $Q$ denote either the kinetic ($K$) or potential ($U$) energy in a block inside a larger system. Average values of $Q$ can be obtained in two ways. From theory, the thermal average (denoted by angled brackets with subscript $th$, $\langle Q \rangle_{th}$) is obtained in standard statistical mechanics by Boltzmann averaging. From simulations, the average is obtained by averaging over all time-steps then taking the arithmetic mean over all blocks in the system (denoted by angled brackets with subscripts $t, bl$, $\langle Q \rangle_{t,bl}$). If the ergodic hypothesis holds, and if simulations and theory yield equivalent results, these two averages will equal each other

$$\langle Q \rangle_{t,bl} = \langle Q \rangle_{th}. \tag{18}$$

The $C \leftrightarrow (\Delta E)^2$ relation is derived in canonical theory by taking the derivative of the thermal average with respect to $kT$ to get a dimensionless heat capacity, then using the product rule of calculus to obtain the variance for the fluctuations

$$\frac{\partial \langle Q \rangle_{th}}{\partial (kT)} = \frac{(\Delta_{th} Q)^2}{(kT)^2} \qquad \left( (\Delta_{th} Q)^2 \equiv \langle Q^2 \rangle_{th} - \langle Q \rangle_{th}^2 \right) \tag{19}$$

Equation (19) applies to any system with thermal average energies that are determined solely by Boltzmann's factor, making this $C \leftrightarrow (\Delta E)^2$ relation a very basic test for thermal behavior. We ask the question: if Eq. (18) connects time-averages to thermal-averages, does Eq. (19) hold for time-averages? Specifically, does the $C \leftrightarrow (\Delta E)^2$ relation apply to time-averages

$$\frac{\partial\langle Q\rangle_{t,bl}}{\partial(kT)} = \frac{\left(\Delta_{t,bl}Q\right)^2}{(kT)^2} \; ? \qquad \left(\left(\Delta_{t,bl}Q\right)^2 \equiv \langle (Q(t) - \langle Q\rangle_t)^2\rangle_{t,bl} \right. \tag{20}$$

MD simulations of several models establish that Eq. (18) is accurately obeyed for average energies, so that the ergodic hypothesis applies to time- and thermal-averages. However, at sufficiently low temperatures all these models show clear deviations from Eq. (20).

*5.2 Thermal Fluctuations and MD Simulations*

MD simulations of the Lennard-Jones (L-J) model were made on systems of 442,368 atoms using parameters for crystalline argon [27,51]. The systems were subdivided into blocks containing 1-2048 atoms to study local thermal behavior. Each block gives excellent agreement with the equipartition theorem for average thermal behavior. One example is shown in Figure 6. From $T < 10$ K, the black line (with error bars) near unity gives $\delta_T\langle U\rangle_{t.bl}/\delta_T\langle K\rangle_{t,bl} = 0.9998 \pm 0.0013$ for the ratio of partial specific heats from potential and kinetic energies. Here, $\delta_T$ denotes the temperature difference, e.g. $\delta_T U \equiv [U(T + \delta T) - U(T)]$. This ratio of partial specific heats is well within statistical uncertainty of unity, indicating that both $U$ and $K$ obey the equipartition of energy.

Symbols in the main part of Figure 6 show the temperature dependence of energy fluctuations for block sizes given in the legend. The blocks are a virtual construct, with an equal number of interacting neighbors for every atom. The $C \leftrightarrow (\Delta E)^2$ relation predicts that energy fluctuations should be proportional to specific heat, whereas the ratio $\langle(\Delta_t U)^2\rangle_{t,bl}/\langle(\Delta_t K)^2\rangle_{t,bl}$ tends to diverge as $T \to 0$, reaching 10 to 100 times the ratio of unity. These results establish that MD fluctuations deviate from canonical-ensemble behavior. The inset of Figure 6 shows why: time-dependent correlations within blocks (black) and between blocks (red) are strongly anticorrelated. Thus, most of the energy in each fluctuation comes from the surrounding shell of neighboring blocks, not from the large heat bath needed for the canonical ensemble. Specifically, the black line and symbols show auto correlations as a function of time, $\Gamma_{0,0}(\tau)/\Gamma_{0,0}(0)$, where $\Gamma_{0,0}(\tau) \equiv \langle U_0(t)U_0(t+\tau)\rangle_{t,bl}$. Due to the normalization $\Gamma_{0,0}(\tau)/\Gamma_{0,0}(0)$ starts at 100%, then exhibits damped oscillations as a function of $\tau$. The red line and symbols show pair correlations between each central block and its six nearest-neighbor blocks, $\Gamma_{0,1}(\tau) \equiv \langle U_0(t)U_1(t+\tau)\rangle_{t,bl}$, which are strongly anticorrelated with the autocorrelations. Indeed, at $\tau = 0$, more than 50% of the energy of the fluctuation comes from the six nearest neighbor blocks. Thus, when the energy from all $3^3 - 1 = 26$ blocks in the local shell of blocks that touch the central block are included, essentially 100% of the initial energy in each fluctuation comes from this local shell, not the large heat bath needed for Boltzmann's factor. A theory for these localized fluctuations based on fast energy modulations yields the dashed lines in the main part of Figure 6 that mimic the simulations with no adjustable parameters [27]. We now outline the theory.

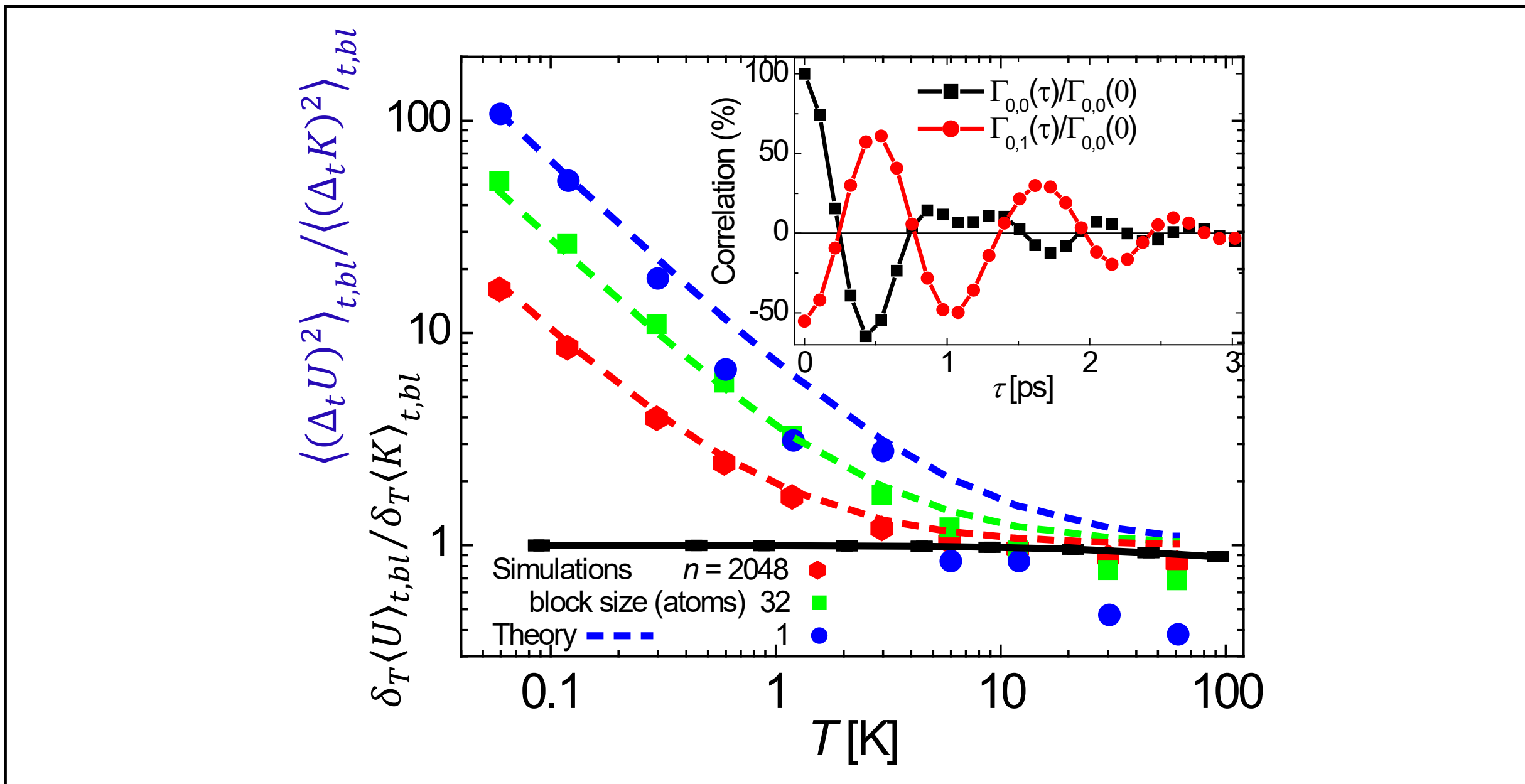

*Figure 6* Ratio of partial specific heats (black line with error bars) and fluctuations (symbols) from MD simulations of the L-J model using parameters for argon. From $T < 10$ K, $\delta_T\langle U\rangle_{t,bl}/\delta_T\langle K\rangle_{t,bl} = 0.9998 \pm 0.0013$ for the ratio of partial specific heats of potential ($U$) and kinetic ($K$) energies. Symbols show that the ratio of their fluctuations tend to diverge as $T \to 0$ for individual atoms (blue) and blocks containing 32 (green) and 2048 (red) atoms. Dashed lines come from a theory for modulated energies, with no adjustable parameters. The inset shows time-dependent autocorrelations (black) and pair correlations (red) between each central block and its six neighboring blocks.

The key assumption in the theory is that fast modulations of interactions between next-nearest-neighbor (nnn) atoms are effectively adiabatic, so that they are time-averaged before thermal averaging. The theory starts with the net potential energy of an atom in a crystal as a function of the displacement ($x$) from its equilibrium position ($x = 0$) and as a function of time ($t$) from ground-state modulations due to the motion of nnn atoms, yielding

$$u(x,t) = ax^2 + bc(t). \tag{21}$$

Referring to Figure 7, the quadratic term ($a$) comes from interactions between nearest-neighbor (nn) atoms, and the linear term $b$ from interactions between nnn atoms. Now, assuming harmonic motion of the nnn atoms, the modulation term can be written as

$$c(t) = -\mathcal{A}[\sin(\omega t + \alpha) + \sin \omega t], \tag{22}$$

For classical MD simulations that have equipartition of energy, the modulation amplitude, $\sqrt{\langle(A_\pm)^2\rangle_{th}} \equiv \mathcal{A}$, is given by the thermal average amplitude

$$\mathcal{A} = \sqrt{k_B T/(2a)}, \tag{23}$$

The main mechanism causing deviations from the $C \leftrightarrow (\Delta E)^2$ relation can be traced to the fact that the time average of Eq. (22) is zero, whereas the time average of $[c(t)]^2$ is not zero. This modifies the expression for potential energy fluctuations [27] so that Eq. (20) becomes

$$\frac{(\bar{\Delta}_{th}U)^2}{(k_BT)^2} = \frac{1}{2} + \frac{1}{2}\frac{b^2}{akT}\,. \tag{24}$$

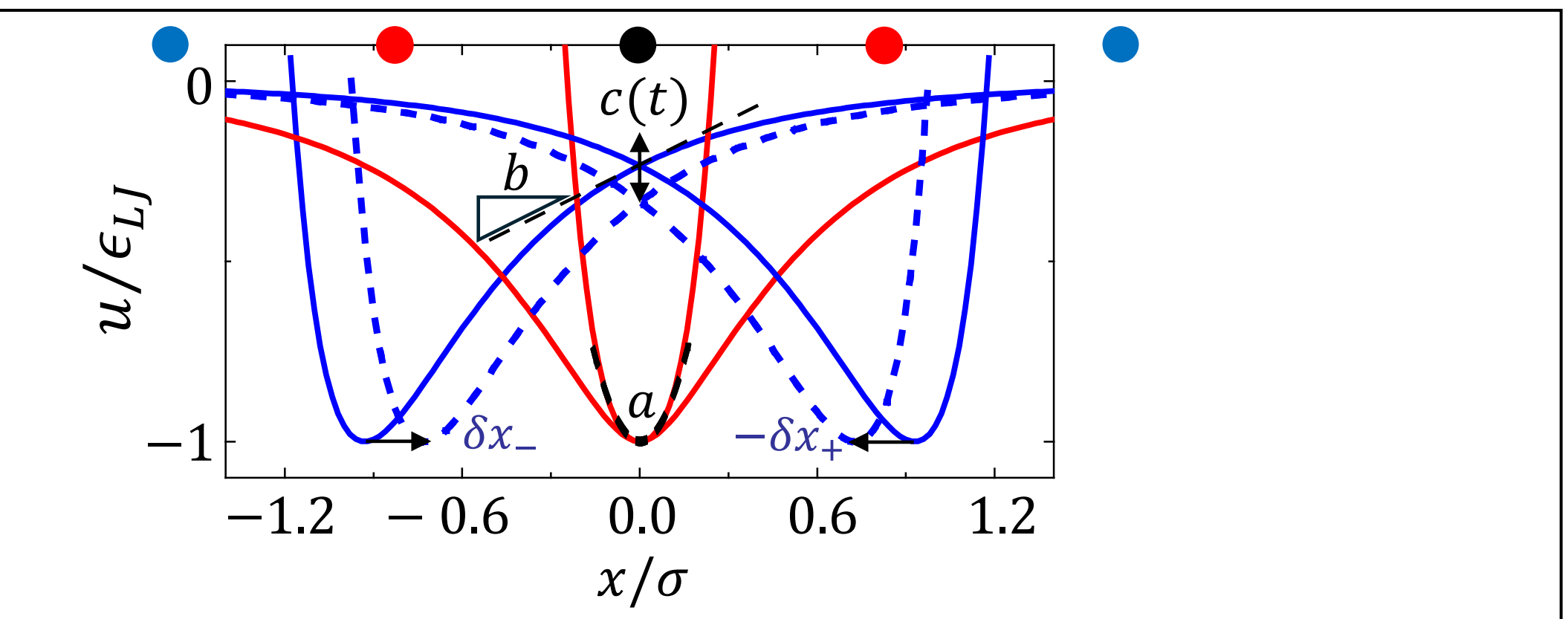

*Figure 7* Circles depict nn atoms (red) and nnn atoms (blue) that contribute most to the energy of the central atom (black). Main figure shows potential energy at the central atom as a function of its displacement along the $X$-axis, normalized by L-J units. Solid lines are from nn atoms (red) and nnn atoms (blue). Dashed blue lines are from a pair of nnn atoms that are displaced by a breathing mode, $\mp\delta x_{\pm}$, showing how the energy of the central atom is modulated, $c(t)$. Dashed black lines show how for the central atom in equilibrium, the potential-energy minimum has a constant curvature ($a$) from nn atoms, with a constant slope ($b$) from each nnn atom.

The interaction parameters for the L-J potential in its ground state fcc lattice are $a = 144\epsilon_{LJ}/(2^{1/3}\sigma^2)$ and $b = -63\epsilon_{LJ}/(2^{14/3}\sigma)$ [28]. For argon atoms, the energy scale is $\epsilon_{LJ}/k_B = 118.2$ K and the length scale is $\sigma = 0.3405$ nm [51]. Utilizing a proposed connection between anomalous diffusion and the block-size dependence for energy fluctuations, the theoretical expression for potential energy fluctuations of a block of $n$ atoms is

$$\frac{(\Delta_{t,bl}U)^2}{(\Delta_{t,bl}K)^2} - 1 = \frac{21^2}{2^{13}}\frac{\epsilon_{LJ}}{k_BT}n^{1/4} \approx 0.053833\frac{118.2}{T[K]}n^{1/4}. \tag{25}$$

The dashed curves in Figure 6 show that Eq. (25) quantitatively characterizes the excess fluctuations. Note the qualitative similarity between simulations having excess fluctuations shown in Figure 6 and measurements having excess specific heat shown in Figure 2. This connection is quantified in Ref. [27]. Thus Eq. (24) replaces the usual $C \leftrightarrow (\Delta E)^2$ relation for systems having fast dynamics that must be time-averaged before thermal-averaging. Similar separation of time scales and isolation from the heat bath are characteristic of measurements of thermodynamic heterogeneity, as described in Section 3, sketched in Figs. 2-3, and deduced from NMR measurements in Figure 4A. From these measurements, localized degrees of freedom often exhibit effective temperatures that differ significantly from the homogeneous heat bath, as found in liquids, glasses, spin glasses, polymers and crystals. Thus, one feature of nanothermodynamics that is crucial to the theory for fluctuations in MD simulations is the ubiquity of thermodynamic heterogeneity. Another feature is that appropriate ensembles must be used to ensure accurate behavior.

## 6. Nanoscale Fluctuations and the Second Law of Thermodynamics

### *6.1 A Creutz-Like Model for Investigating Nanoscale Fluctuations*

Nanothermodynamics impacts several aspects of nanoscale fluctuations. Two consequences of including finite-size thermal effects are that both the choice of ensemble [52] and the choice of dynamics [17] influence the fluctuations. Thus, if all sources of energy from finite-size thermal effects are included, fluctuations exhibit features that are not contained in standard theories for thermal fluctuations that are based on microscopic reversibility and ensemble equivalence [53].

A statement of the second law of thermodynamics is that the entropy of a closed system will most likely increase monotonically until it reaches its maximum value, which defines the thermal equilibrium. Thus, investigations of the 2nd law are ideally done using closed systems, having an explicit heat bath that thermalizes the system with no external influences. Perhaps the simplest such model, adapted [17] from the work of Creutz [16], involves 1-D chains of Ising spins coupled to a bath of Einstein oscillators (that Creutz called "demons"). This simplified 1-D Creutz model was first used to study how the 2nd law applies not only to large and complex systems, but also to small and simple systems [17]. Sketches of the model are shown on the right-hand side of Figure 5 (above) and repeated in Figure 8B (below). An advantage of this model is that the exact entropy of the system and its bath can be calculated at every step. Moreover, analytic expressions for the stable equilibrium of the 1-D Ising model are known, as given in section 4.2 (above). Here we will show that, at least for this model, the 2nd law requires an intrinsically irreversible microscopic step, providing a clear counterexample to the usual assumption that large systems have thermal behavior that emerges from intrinsically reversible microscopic dynamics. Another result shown by this model is that fluctuations in total entropy are best described by the 2nd law of thermodynamics, not by standard statistical theories based on Boltzmann's factor.

### *6.2 Reversible and Irreversible Simulations of the 1-D Creutz Model*

The simulations shown in Figure 8 come from a system of $N = 250{,}000$ spins with an equal number of demons. The closed system contained a fixed amount of energy ($1/2$ unit per spin), shared between the spins and demons, yielding an average temperature of $kT/J = 1.30$. Again, a crucial feature of the 1-D Creutz model is that the exact entropy of the spin system and its bath can be calculated at every step. The solid lines in Figure 8 show the total entropy (from spins and demons, $S_t$) per spin as a function of time from simulations with two types of dynamics: reversible (black) and irreversible (red). Both simulations evolved from the same initial spin configuration that was highly ordered (low entropy). Details about the different dynamics are given in [17]. Briefly, reversible dynamics utilizes three randomly generated (but fixed) arrays: one array sets the sequence of spins to be updated while the other two arrays govern which two demons couple to each spin. In contrast, irreversible dynamics utilizes a new random number for each of these steps.

Figure 8A shows that at the start of the simulations, $S_t$ tends to increase with time, as expected from the 2nd law of thermodynamics for closed systems that are initially out of equilibrium. However, the nature of this increase is strongly dependent on the dynamics. Irreversible dynamics (red) yields a rapid and mostly monotonic increase, while reversible dynamics (black) shows slower relaxation with large oscillations. Figure 8C shows that at the end of the simulations the differences are even more pronounced. Of course, $S_t$ from irreversible dynamics (red) stays at or near its maximum ($S_m$), as expected from the 2nd law. However, for reversible dynamics (black) the simulation exhibits exact reversibility, with every step down in entropy that was taken up in entropy at the start, a clear demonstration of Loschmidt's paradox for dynamics that is reversed at the midpoint of its evolution.

Most standard resolutions of Loschmidt's paradox rely on the fact that for large and complex systems it is difficult to exactly reverse the dynamics at the midpoint of an evolution. In other words, a large isolated system has a vanishingly small fraction of initial conditions that cause $S_t$ to systematically decrease. However, as shown in Figure 8, it is relatively easy to find special sets of initial conditions for reversible systems [54] of any size by simulating from an initially low-entropy state before reversing the dynamics. More importantly, there are fundamental reasons why the standard argument about initial conditions cannot explain the 2nd law, at least for the 1-D Creutz model. First, from Figure 8B, $S_t$ from irreversible dynamics remains significantly higher than $S_t$ from reversible dynamics, regardless of the initial state. Second, as shown in figure 2A of [17], after an irreversible simulation $S_t$ always decreases when the dynamics is switched to reversible. Finally, the difference in entropy density between reversible and irreversible dynamics increases with decreasing system size, as shown in figure 3 of [17]. Thus, the usual assumption that only special initial conditions cause entropy to decrease in reversible systems cannot solve Loschmidt's paradox, at least not for the 1-D Creutz model.

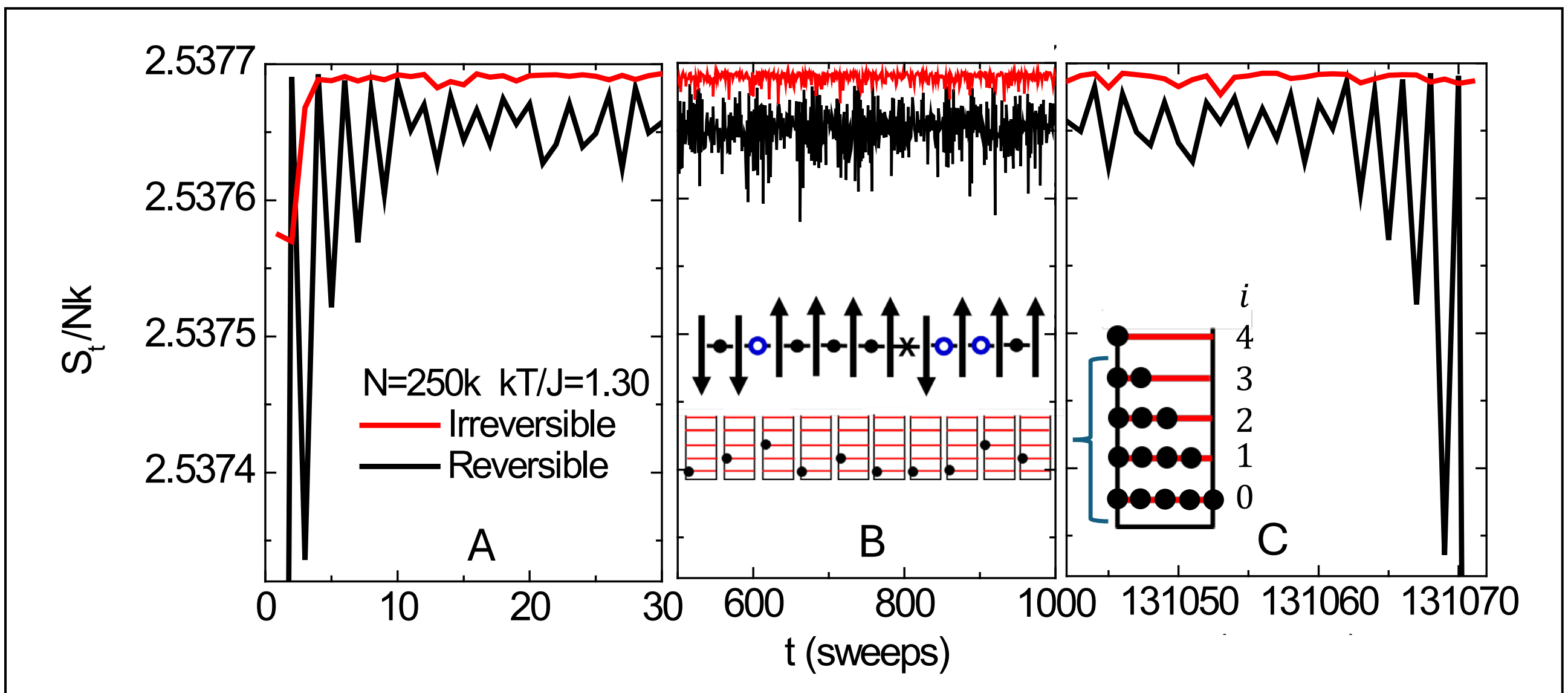

*Figure 8* Results from simulations of a 1-D Creutz model sketched in (B). It is a closed system containing $N = 250{,}000$ Ising-like spins coupled to a heat bath of $N$ Einstein oscillators ("demons") that allows exact calculation of total entropy at every step. The Ising spins form finite chains by having breaks in their interactions (open circles in the sketch), as needed for stable equilibrium. Solid lines in the graphs show total entropy per particle as a function of time at the beginning (A), middle (B), and end (C) of the simulations with dynamics that is intrinsically reversible (black) or irreversible (red). (All graphs have the same vertical scale, but different horizontal scales.) The dynamics is reversed at the midpoint of the simulation (not shown). Note how reversible dynamics is exactly reversible, following every step down in entropy that was taken up in entropy at the start, even after more than $6.5x10^7$ steps, demonstrating Loschmidt's paradox and violations of the 2nd law of thermodynamics. In (B), note how reversible dynamics shows nearly normal fluctuations about an average value, consistent with standard theories for fluctuations, whereas irreversible dynamics has $S_t$ that tends to have one-sided fluctuations, decreasing from a maximum entropy, consistent with the 2nd law. The sketch in (C) indicates how the relative occupation of energy levels ($i$) can be used to extract local temperatures.

*6.3 Comparison to Standard Theories for Fluctuations*

Most theories of equilibrium fluctuations predict Gaussian distributions for thermal quantities. This, of course, comes from the central limit theorem for a large number of uncorrelated, randomly distributed quantities with a finite variance. Many derivations start by expanding the entropy change as a function of the quantity. Specifically, for fluctuations in energy, $\Delta E$, assuming analyticity, from an expansion of the change in entropy, $p(\Delta E) \propto \exp(\Delta S/k) \approx \exp\left[(\Delta E/k)(\partial S/\partial E) + \frac{1}{2}(\Delta E/k)^2(\partial^2 S/\partial E^2)\right]$. Thermodynamic identities yield $\partial S/\partial E = 1/T$ and $\partial^2 S/\partial E^2 = -1/(CT^2)$. Linear changes in energy in systems of any size are accommodated by the heat bath (subscript $b$), $\Delta E_b = -\Delta E$, so that the first term in the expansion becomes Boltzmann's factor [55]. The second term yields a Gaussian distribution for equilibrium energy fluctuations. This Gaussian distribution has many applications, including characterizing the slow equilibrium dynamics of supercooled liquids [14].

Fluctuation theorems have extended the range for theoretical treatments of thermal fluctuations to non-equilibrium conditions. Various fluctuation theorems are described in [53,56,57]. Applications of the theorems to experiments include characterizing the free energy change while stretching RNA [58] and the fluctuations of colloidal particles [59,60]. Of particular relevance for analyzing the 1-D Creutz model are detailed fluctuation theorems that apply to the total entropy of a closed system [18-21]. In general, these theorems yield the probability that the total entropy increases (positive entropy production) divided by the probability that it decreases (negative entropy production), as given by

$$p(\Delta S_t)/p(-\Delta S_t) = \exp(\alpha \Delta S_t) \tag{26}$$

Here, $\alpha = 1$ for a driven system measured over a time interval [18] and along various stochastic trajectories [19,21]. Alternatively, $\alpha$ depends on a diffusion constant, $D$, and in general on the nonlinearity of the force [20], yielding $\alpha \neq 1$. Note that Gaussian distributions invariably obey Eq. (26). Specifically, using $p(\Delta S_t) \propto \exp[-(\Delta S_t - B)^2/\omega^2]$ gives

$$p(\Delta S_t)/p(-\Delta S_t) = \exp(4B\Delta S_t/\omega^2) \tag{27}$$

in agreement with Eq. (26) if $\alpha = 4B/\omega^2$. A remarkable feature of fluctuation theorems (that will not be addressed here) is that they may also apply to non-equilibrium conditions.

To compare theoretical treatments of fluctuations to simulations of the 1-D Creutz model it is useful to use histograms that convert entropy as a function of time from Figure 8 to distributions of multiplicity as a function of entropy, Figure 9. Two distinct types of distributions are found. Reversible dynamics (filled symbols) exhibit distributions that are approximately Gaussian, consistent with the central limit theorem when $S_t$ can fluctuate freely up and down. However, with decreasing system size the Gaussians become increasingly skewed towards low entropy, opposite to standard fluctuation theorems that favor high-entropy states. From Figure 9 It is evident that this asymmetry increases with decreasing system size because larger fluctuations allow $S_t$ to approach the maximum entropy ($S_m$) shown by the irreversible dynamics (open symbols). Indeed, this $S_m$ sets an effective upper limit for fluctuations in $S_t$. More importantly, equilibrium (irreversible) simulations have only asymmetric fluctuations, yielding one-sided distributions that are characterized by

$$p(S_t) = A\exp(S_t/k) \qquad (S_t \leq S_m). \tag{28}$$

Here, the normalization constant for a continuous distribution is $A = 1/[\exp(S_m/k) - 1]$. The solid lines in Figure 9 show good agreement between Eq. (28) and the open symbols from irreversible dynamics. Equation (28) is consistent with Einstein's reversal of Boltzmann's relation that he used to describe equilibrium fluctuations [61]. If $\alpha = 2$, Eq. (28) may also be consistent with Eq. (26), except at the limit of $S_t = S_m$ where entropy can only decrease.

The asymmetry in the distribution of $S_t$ during irreversible dynamics gives insight into the physics of equilibrium thermal fluctuations. Standard fluctuation theories predict that entropy increases are more likely than entropy decreases. Equation (28) adheres to this prediction for most values of $S_t$, except at $S_m$ where $S_t$ can only decrease. This maximum in entropy is evident from the one-sided fluctuations shown by irreversible dynamics (red line) in Figure 8 B, and by the one-sided distributions in Figure 9. Although the one-sided distributions differ from usual predictions of standard statistical mechanics, this asymmetry is needed to yield maximum entropy at the highest multiplicity, another indication that only irreversible dynamics is consistent with the 2nd law of thermodynamics.

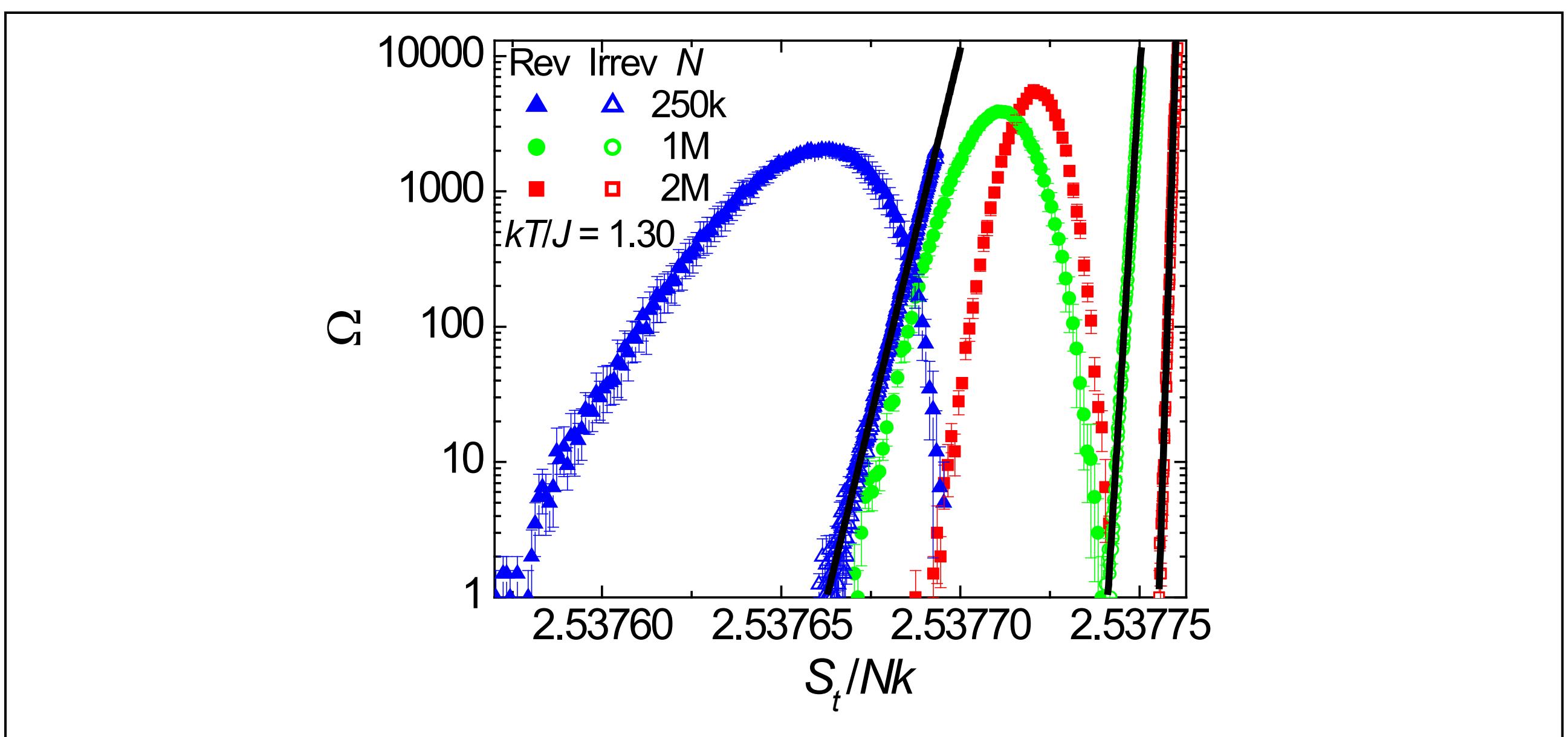

*Figure 9* Histograms of multiplicity (on a logarithmic scale) as a function of total entropy per particle. Symbols (with error bars) show results from simulations on three system sizes (given in the legend) and two types of dynamics: reversible (filled) or irreversible (open). Reversible dynamics shows Gaussian-like behavior, qualitatively consistent with standard theories for fluctuations. However, the distributions are skewed by the maximum entropy for smaller systems. Irreversible dynamics has a multiplicity that increases exponentially with increasing entropy, with the maximum multiplicity at maximum entropy, consistent with the 2nd law of thermodynamics. Solid lines show fits to the irreversible multiplicities using Eq. 28 with $A$ as the only parameter.

### *6.4 Effective Temperatures from the 1-D Creutz Model*

Because even von Neumann recognized that "nobody really knows what entropy is" [62], it is useful to go beyond the concept of maximum entropy and the 2nd law to further justify the need for irreversible dynamics. An appealing feature of the 1-D Creutz model is that it utilizes an explicit heat bath of demons to yield local temperature as an output of the simulation, not as an input, as indicated by the sketch in Figure 8C. One application is to use the distribution of energies in each oscillator for location-specific temperatures that are ideal for characterizing the thermodynamic heterogeneity measured by TRSH and NSHB described in Section 3. Another application is to use the relative occupation of adjacent energy levels to obtain an energy-specific temperature that is ideal for assessing the accuracy of the Boltzmann distribution. Indeed, because the demons that form the thermal bath have equal energy spacing, $J$, Boltzmann's factor gives the relative probability of energy level $i$ as, $p_i \propto \exp(-iJ/kT)$. Thus, an energy-specific effective temperature ($T_i$) between adjacent levels is given by

$$kT_i/J = 1/\ln(p_i/p_{i+1}). \quad (30)$$

Figure 10 shows the energy-level dependence of this effective temperature from simulations of the 1-D Creutz model with three system sizes and both types of dynamics. Irreversible dynamics gives a single well-defined temperature for all levels. Reversible dynamics gives an effective temperature that oscillates between even and odd levels. Again, reversible dynamics differs significantly from expected thermal behavior. At the very least, reversible dynamics yields a non-equilibrium steady state condition between the system and its bath, even when there are no external forces. Indeed, as shown in figure 2 of [17], it is this failure of the heat bath to follow Boltzmann statistics that is the main cause of reduced $S_t$ for reversible dynamics, as shown here in Figs. (8) and (9).

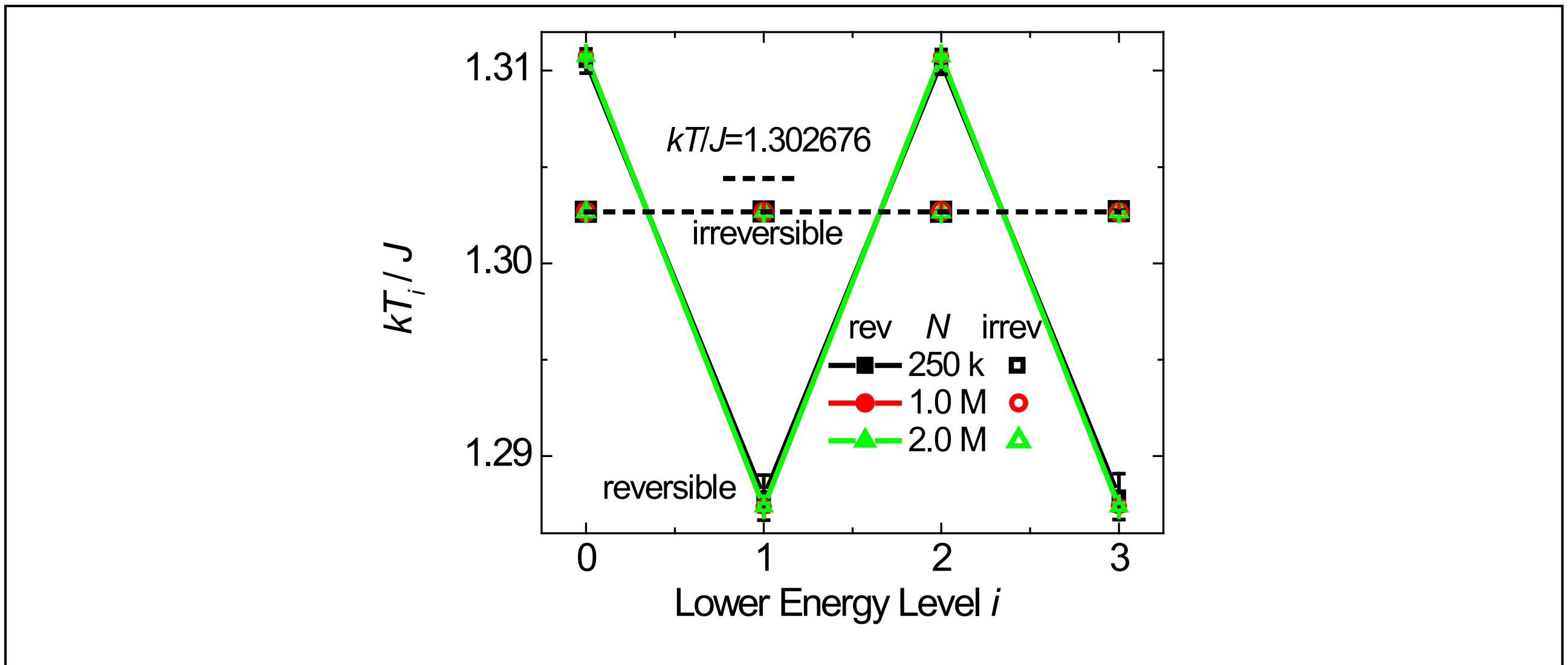

*Figure 10* Effective temperatures as a function of energy level using Eq. (30), from simulations on systems with sizes given in the legend. Reversible dynamics yields temperatures that differ by about 2% between even and odd levels, which is about 20 standard deviations as given by the error bars. Irreversible dynamics shows a single well-defined temperature.

### *6.5 Conclusions from Simulations of the 1-D Creutz Model*

Three questions arise regarding the behavior shown in Figs. 8-10: what is the role of nanothermodynamics, can the simplistic 1-D Creutz model be connected to real-world systems, and what might be the source of intrinsically irreversible dynamics in nature? For the first question, all results presented here yielding maximum entropy require that the Ising model be able to access the stable thermal equilibrium of the nanocanonical ensemble. In fact, if the chain of Ising spins is constrained to be homogenous (depicted by the top sketch in Figure 5) then there is negligible difference between reversible and irreversible dynamics. Furthermore, total entropy is never maximized so that the distributions are not one-sided, and there are no odd energy levels with different effective temperatures for Figure 10. Thus, the stable thermal equilibrium of nanothermodynamics is essential for studying the influence of the 2$^{nd}$ law on fluctuations.

An overall answer to the second question is that Creutz demons play a crucial role in connecting to physical reality, which cannot be matched by any idealized heat bath such as the Metropolis algorithm or Langevin equation. On a practical level, random numbers are an inherent feature of idealized heat baths so that their simulations cannot be made reversible.

Furthermore, as shown in figure 2 of [17], entropy of the spin system *decreases* for irreversible dynamics: the increase in $S_t$ needed for the 2nd law (shown here in Figs. 8 and 9) requires the explicit heat bath. Similarly, an explicit heat bath and maximum entropy are also needed for a consistent local temperature that is not externally imposed on the system, as shown in Figure 10. On a fundamental level, the demons ensure that local energy is always conserved, unlike idealized heat baths where changes in local energy are assumed to be instantaneously absorbed by an infinite bath. This fundamental aspect also connects the 1-D Creutz model to MD simulations (Section 5) because both types of simulations are done in the most-basic microcanonical ensemble. Indeed, conservation of local energy is evident in the insert of Figure 6 showing that potential energy is traded back-and-forth between each central block and its surrounding shell of neighboring blocks. It is these local oscillations that insulate fast degrees of freedom from the heat bath, preventing them from having a well-defined temperature. Similarly, the oscillations in $S_t$ from the 1-D Creutz model with reversible dynamics (clearly visible in Figs. 8A and 8C) can be attributed to energy that is repeatedly traded back-and-forth between each spin and its heat bath of demons. This interpretation is consistent with the observation that the oscillation amplitude decreases with increasing number of demons that connect to each spin, as shown in Fig. 1 of reference [17]. Whereas, for irreversible dynamics, the random connection between spins and multiple demons efficiently disperses the initially localized energies. In contrast, reversible dynamics causes all simulations to exhibit persistent oscillations that inhibit the systems from reaching states of maximum entropy. Thus, the evidence indicates that an explicit heat bath more-accurately represents reality than idealized heat baths, and that intrinsically irreversible dynamics is a common requirement of any closed system that is to obey the 2nd law of thermodynamics.

So far there is no definitive answer to the third question. Although irreversible simulations of the 1-D Creutz model utilize a pseudo-random number generator, what is the source of randomness in real systems? In reference [17] it is speculated that randomness in nature could come from the quantum measurement process. Specifically, if the process connecting the system to its bath is irreversible, then the precise bath particle to be used is unknown until the connection is made. This connection is particularly impactful if the bath contains indistinguishable particles that become distinguishable when the measurement is made. Indeed, it is the nature of this connection between system and bath that is the only difference between reversible and irreversible simulations shown in Figs. 8-10. Presumably, it is the fact that random connections have higher entropy than reversible connections that allows the system to reach its maximum entropy. Future work will be necessary to understand how the connection between a system and bath controls the behavior of the combined system.

## VII. Summary

In this review it is emphasized that large systems usually favor subdividing into stable distributions of internal subsystems, which is the basis of nanothermodynamics. In section 2, the history, foundations, and significance of nanothermodynamics are discussed. Figure 1 summarizes the progression of the fundamental equation of thermodynamics, culminating in Hill's subdivision potential, $\mathcal{E}$, that is needed for the systematic treatment of finite-size thermal effects.

In section 3, experimental evidence is given for thermodynamic heterogeneity inside bulk materials due to a distribution of effective local temperatures, $T_i$. The evidence for distinct values of $T_i$ during the response of most types of materials – including liquids, glasses, spin glasses, polymers, and crystals – justifies the need for nanothermodynamics. Figure 2 shows results from measurements of $\Delta T_i$ as a function of time on high-purity single crystals, yielding equilibrium specific heats that significantly exceed Debye's theory. The essentially inverse process of nonresonant spectral hole burning establishes that there is negligible correlation between the slow and fast degrees of freedom in a spin glass (Figure 3) and a glass-forming liquid (Fig 4). Furthermore, Figure 3 shows that $T_i$ may exceed the bath temperature by 2 K, or more, heating a significant fraction of the sample into the paramagnet phase, while the rest of the sample remains a spin glass. Figure 4A is a picture that emerges from measurements of nuclear magnetic resonance indicating that the thermodynamic heterogeneity in glass-forming liquids occurs on length scales of nanometers.

Section 4 outlines the need for nanothermodynamics to solve two standard models in statistical physics. First, the stable thermal equilibrium of Ising's original model for finite chains of spins is derived two ways, using two different ensembles starting from two different Hamiltonians. It is emphasized that standard thermodynamics and statistical mechanics give no guidance on how to find this equilibrium in the nanocanonical ensemble. Second, the stable equilibrium of the semi-classical ideal gas is described, which is needed to ensure that entropy is extensive for systems of any size. This novel solution to Gibbs' paradox may also explain the lack of isotope effects in measured entropies of dilute gases. Figure 5 shows how minimizing the total free energy requires subdividing large systems into the stable distribution of small subsystems, which is the basis of nanothermodynamics.

In section 5, MD simulations are used to investigate the excess energy fluctuations found at low temperatures (Figure 6) that are qualitatively similar to the excess specific heat measured in crystals (Figure 2). A theory is described based on fast (adiabatic) modulation of ground state energies due to localized fluctuations, as sketched in Figure 7. This theory is consistent with the anticorrelated local energies in the simulations (inset of Figure 6), and with measurements of thermodynamic heterogeneity. The theory gives quantitative agreement with the excess fluctuations, and a correction to the fluctuation relation for degrees of freedom that are too fast to fully couple to the heat bath, Eq. (24).

In section 6, simulations of a 1-D Creutz model are used to investigate nanoscale thermal fluctuations and the 2nd law of thermodynamics. This model is comprised of chains of Ising-like spins coupled to an explicit heat bath of Einstein oscillators. The model has several appealing features: the total energy is always exactly conserved, the spins can reach their stable thermal equilibrium, the heat bath yields the distribution of local temperatures, the dynamics can be made fully reversible, and the total entropy ($S_t$) can be calculated exactly at every step. As shown in Figure 8, maximizing $S_t$ requires an intrinsically irreversible microscopic step, providing a counterexample to the usual claim that thermal equilibrium can emerge from reversible dynamics. Figure 10 shows that this irreversible step is also needed for a well-defined temperature. Figure 9 shows that the distribution of $S_t$ from fluctuations is strongly dependent on the dynamics. For thermodynamically consistent behavior, the fluctuations are best described by Einstein's reversal of Boltzmann's relation and the 2nd-law, not by any recent fluctuation theorem. Section 6 finishes with a discussion about the relevance of the 1-D Creutz model to real systems, and a possible source of intrinsic irreversibility in nature.

In conclusion, many measurements establish the necessity of nanothermodynamics for the thermal and dynamic behavior of most types of materials. Nanothermodynamics is crucial in finding the stable thermal equilibrium of standard models from statistical physics. Deviations of MD simulations from a fluctuation relation can be quantitatively characterized by including fast degrees of freedom from localized modes that are effectively decoupled from the heat bath. Nanoscale fluctuations in a simple model for a closed system are best described by Einstein's reversal of Boltzmann's relation and the 2nd law of thermodynamics, not by recent fluctuation theorems. It is anticipated that the stable thermal equilibrium introduced by Hill 62 years ago, and its adaptation to nanothermodynamics, will continue to find applications that give new insights into experiments, theory, and simulations.

**Acknowledgments**

I have benefited greatly from the generosity and wisdom of Sumiyoshi Abe.

**Funding**

This research received no external funding.

**Declaration of Competing Interest**

The author declares that he has no known competing financial interests or personal relationships that could have appeared to influence the work reported in this paper.